\documentclass[conference]{IEEEtran}
\usepackage{cite}
\usepackage{amsmath,amssymb,amsfonts}
\usepackage{algorithm} 
\usepackage{algorithmic}
\usepackage{graphicx}
\usepackage{textcomp}
\usepackage{xcolor}
\usepackage{fancyhdr}
\usepackage[hyphens]{url}

\usepackage{tikz}

\usepackage{svg}
\usepackage[normalem]{ulem}
\usepackage{titling}
\usepackage{wrapfig}
\usepackage{comment}
\usepackage{xspace}
\usepackage{listings}
\usepackage{cite} 
\usepackage{multicol}
\usepackage{tikz}
\usepackage[export]{adjustbox}
\usepackage{wrapfig}
\usepackage{enumitem}
\def\BibTeX{{\rm B\kern-.05em{\sc i\kern-.025em b}\kern-.08em
    T\kern-.1667em\lower.7ex\hbox{E}\kern-.125emX}}

\usepackage{caption}%
\usepackage{bm}

\newcommand{\red}[1]{\textcolor{black}{#1}}
\newcommand{\rev}[1]{\textcolor{black}{#1}}
\newcommand{\blue}[1]{\textcolor{black}{#1}}
\newcommand{\projectname}{IntersectX\xspace}

\newcommand{\AvgSpeedupGRMER}{40.1$\times$}
\newcommand{\MaxSpeedupGRMER}{181.8$\times$}
\newcommand{\AvgSpeedupCPU}{ 10.7$\times$}
\newcommand{\MaxSpeedupCPU}{ 83.9$\times$}

\newcommand{\gramerFourCliqueAvgSpeedup}{ 85.5$\times$}
\newcommand{\gramerFiveCliqueAvgSpeedup}{ 121.3$\times$}
\newcommand{\gramerTriAvgSpeedup}{ 32$\times$}

\newcommand{\nestSpeedUp}{ 1.357$\times$}
\newcommand{\baselineCPU}{\blue{AutoMine }}

\lstdefinestyle{MyScala}{
  numbers=left,  
  firstnumber=1,
  numberfirstline=true,
  numberstyle=\tiny\ttfamily,
  language=scala,
  aboveskip=3mm,
  belowskip=3mm,
  showstringspaces=false,
  basicstyle={\scriptsize\ttfamily},
  keywordstyle=\bfseries,
  captionpos=b,
  columns=flexible,
  xleftmargin=0.05\textwidth, xrightmargin=0\textwidth,
  breaklines=true,
  breakatwhitespace=true,
  mathescape=true,
  tabsize=4
}
\title{\huge \vspace{-10mm} \projectname: An Efficient Accelerator for Graph Mining }

\date{}

\usepackage{authblk}

\author[1]{Gengyu Rao}
\author[1]{Jingji Chen}
\author[1]{Jason Yik}
\author[1]{Xuehai Qian}
\affil[1]{University of Southern California, Los Angeles, CA}

\begin{document}
\maketitle
\pagestyle{plain}

\begin{abstract}
Graph pattern mining applications try to find all embeddings that match specific patterns.
\red{Compared to the traditional graph computation, graph mining applications are computation-intensive.}
The state-of-the-art method, {\em pattern enumeration}, constructs
the embeddings that match the pattern.
The key operation---{\em intersection}---of
two edge lists, poses challenges to conventional architectures and 
\red{requires}
substantial execution time. 

In this paper, we propose {\em \projectname}, a {\em vertically} designed accelerator
for pattern enumeration with 
stream instruction set extension and 
architectural supports based on 
{\em conventional} processor. 
The stream based ISA can be
considered as a natural extension to 
the traditional instructions \red{that}
operate on scalar values.
We develop the \projectname architecture composed of specialized mechanisms that efficiently
implement the stream ISA extensions, 
including:
(1) Stream Mapping Table (SMT) that 
records the mapping between stream ID and stream register;
(2) the read-only Stream Cache (S-Cache) that 
enables efficient stream data movements;
(3) tracking the dependency between streams
with a property of intersection;
(4) Stream Value Processing Unit (SVPU)
that implements sparse value computations;
and (5) the nested intersection translator
that generates micro-op sequences 
for implementing nested intersections.
We implement \projectname ISA and 
architecture on zSim,
\red{and test it with}
seven popular graph mining applications ({triangle/three-chain/tailed-traingle counting, 3-motif mining, 4/5-clique counting, and FSM}) on {ten} real
graphs.
\rev{We develop our own implementation of \baselineCPU  \blue{(InHouseAutomine)}}~\footnote{The codes of ~\cite{mawhirter2019automine} are not 
open source, \blue{we implemented InHouseAutomine based on their paper with other optimizations and achieved comparable performance.} }.
The results show that \projectname significantly outperforms \blue{InHouseAutomine}  on CPU, 
on average \AvgSpeedupCPU and up to 
\MaxSpeedupCPU; and GRAMER, a state-of-the-art graph pattern mining accelerator, based
on exhaustive check, on average \AvgSpeedupGRMER and up to 
\MaxSpeedupGRMER.

\end{abstract}

\section{Introduction}

Graph processing, which attempts to extract the underlying unstructured information of 
massive graph data, has attracted significant attention in 
the recent decade~\cite{ribeiro2019survey,shaw1999methods,uddin2013dyad,duma2014network}. 
Graph computation and graph pattern mining (GPM)
are two major workloads of graph processing~\cite{teixeira2015arabesque}.
Different from the traditional iterative graph computation (e.g., PageRank, BFS, SSSP, etc.)
with simple computations,
GPM applications are {\em computation-intensive}~\cite{teixeira2015arabesque,wang2018rstream,mawhirter2019automine,jamshidi2020peregrine,chen2019pangolin,iyer2018asap}. The goal of GPM is to find all embeddings that match specific patterns~\footnote{\rev{
While the term ``graph mining'' is
used in some papers such as PEGASUS~\cite{kang2009pegasus},
the system in fact performs typical 
graph computation, such as computing the diameter of the graph, the radius of each node and finding 
the connected components. They are not our 
focus and not accelerated by \projectname.} }.
The tasks are 
more challenging since the number of embeddings could be large. For example, \red{with }
WikiVote, a small graph with merely 7k vertices, the number of vertex-induced 5-chain embeddings can reach 71 billion. 

\rev{Accelerating the performance of 
a specific application needs to consider two 
aspects: {\em memory efficiency and computation efficiency}}. The graph computation is 
typically expressed in 
the ``think like a vertex'' (TLV) model~\cite{malewicz2010pregel}
in the graph processing systems~\cite{nguyen2013lightweight,shun2013ligra,gonzalez2012powergraph,malewicz2010pregel,low2014graphlab,dathathri2018gluon,zhu2016gemini} and 
architectures~\cite{ahn2015scalable,zhuo2019graphq,zhang2018graphp,song2018graphr,graphh}.
In the iterative graph computation, 
while the computation that generates the compute 
array updates is typically lightweight, 
these updates incur random accesses.
Thus, the main challenge is the inefficient 
memory access due to 
poor locality and high memory bandwidth
consumption.
The accelerators mainly 
focus on 
hiding communication latency~\cite{zhuo2019graphq},
reducing data movements with 
Processing-In-Memory (PIM) architecture~\cite{zhuo2019graphq,zhang2018graphp,song2018graphr,graphh}, and 
acceleration of 
asynchronous~\cite{yang2020graphabcd,ozdal2016energy,9138946} and iterative~\cite{rahman2020graphpulse} graph
processing.
Before discussing the architectural 
implications of GPM, 
let us consider the two major 
GPM methods. 

The first method is {\em exhaustive check}:
the algorithm enumerates all subgraphs 
with size up to the pattern size---
\red{regardless of}
the specific patterns. 
During \red{the} subgraph expansion process, some 
infeasible 
\red{combinations }
that cannot match the whole
pattern can be pruned early. 
When the subgraphs reach the pattern size,
isomorphic~\footnote{\rev{Two graphs 
$G_0=(V_0,E_0)$ and $G_1=(V_1,E_1)$ are isomorphic iff. there exists and one-to-one mapping $f: V_0 
\rightarrow V_1$ such that $(u,v) \in E_0 \Longleftrightarrow (f(u),f(v)) \in E_1$.}} check is performed between
each candidate
and the pattern graph, if
passed, a valid embedding is identified.
It is used in the \rev{early} graph mining
system Arabesque~\cite{teixeira2015arabesque}. 
The second method is {\em pattern enumeration},
which specifically generates the embeddings
that satisfy the pattern {\em by construction}.
It avoids the expensive isomorphic check
and does not generate infeasible subgraphs.
This method is adopted by recent systems, \red{e.g.},
AutoMine~\cite{mawhirter2019automine}, GraphZero~\cite{mawhirter2019graphzero},
GraphPi~\cite{shi2020graphpi} and
Peregrine~\cite{jamshidi2020peregrine}, which 
achieved significant speedups over Arabesque.
RStream~\cite{wang2018rstream} is a single-machine system that allows users to express patterns using 
relational algebra, so that the 
runtime engine can perform efficient 
tuple streaming.
It lies in between the two method:
the join operation does not precisely construct patterns and isomorphic checking is still needed, however; not all 
subgraphs of the pattern size are enumerated.

\rev{For GPM, the essential memory access 
pattern is {\em edge list access}.
For exhaustive check, they are incurred 
in subgraph expansion~\cite{yao2020locality}.
The main source of intensive computation is
the expensive isomorphic check.}
For pattern enumeration, the key 
operation is 
the {\em intersection between two edge lists}, 
which can construct embededings based on the pattern.
\rev{For example, for two connected
vertices $v_1$ and $v_2$, performing the intersection of the two edge lists will
identify the triangles ($v_1$, $v_2$, $v$), 
where $v$ is the neighbor of both $v_1$ and $v_2$.
During the 
execution of pattern enumeration,
the edge list accesses (memory) and 
the intersection of edge lists (compute)
\red{are performed} alternatively.}

This paper focuses on 
pattern enumeration since is significantly
faster than exhaustive check. The first graph mining accelerator
GRAMER~\cite{yao2020locality} provides
the architectural supports for exhaustive
check. In Section~\ref{sec:intersect_exist}, 
we show that the pattern enumeration on an 
unmodified CPU is likely to run faster
than \red{the} accelerated exhaustive check with GRAMER. 
It shows the importance of developing \red{a}
specialized architecture for the 
state-of-the-art method. 
\rev{For pattern enumeration, {\em we focus on the 
computation efficiency (intersection) rather than
memory efficiency for two reasons}. 
First, the edge list access incurs 
just {\em two random accesses}:
one to get the pointer to
the start of the list and
\red{the other}
to access it.
This is much less than the vertex updates
in \red{the} iterative graph computation---{\em one random access
for each edge}. 
The majority of accesses---the traversal of all neighbors in the list---are sequential. }
Second, we show that
the conventional processor cannot execute
the intersection efficiently (Section~\ref{sec:arch_problem}).
Thus,
{\em the more expensive computation becomes
the bottleneck for GPM.}
The current architectures
that optimize intersection are either designed for 
tensor computations~\cite{hegde2019extensor,gondimalla2019Sparten}, or \red{are} infeasible to execute
the complex code patterns~\cite{dadu2019towards, wang2019stream} in 
GPM, such as computation reuse
and symmetric breaking~\cite{mawhirter2019automine,mawhirter2019graphzero,shi2020graphpi},
without significant efforts. 
Section~\ref{sec:intersect_exist} will 
discuss the state-of-the-art GPM algorithm
and optimizations.

With the understanding of 
the architectural challenge\red{s} and requirement\red{s} of pattern enumeration
based GPM, this paper proposes {\em \projectname}, a {\em vertically} designed accelerator
based on 
{\em conventional} processor 
to accelerate GPM
by making streams first-class citizens in the 
ISA. 
We define a sparse vector 
as a {\em stream},
which can be a {\em key} or 
{\em (key,value)} stream.
Our novel stream ISA extension {\em intrinsically operates
on streams}, realizing both data movement and computation. 
It can be 
considered as a natural extension to 
the traditional instructions 
for ordinary scalar values.
The \projectname architecture is composed of specialized mechanisms that efficiently
implement the stream ISA extensions, 
including:
(1) Stream Mapping Table (SMT) that 
records the mapping between \red{the} stream ID and \red{the} stream register;
(2) the read-only Stream Cache (S-Cache) that 
enables efficient stream data movements;
(3) tracking the dependency between streams
with a property of intersection;
(4) Stream Value Processing Unit (SVPU)
that implements sparse value computations;
and (5) the nested intersection translator
that generates micro-op sequences 
for implementing nested intersections.

To provide good programmability,
similar to Automine~\cite{mawhirter2019automine},
we also provide a GPM compiler to generate stream ISA based GPM implementation. 
Thus, users do not need to write any assembly codes.
Our compiler takes one or multiple patterns as input, 
synthesizes the pattern enumeration 
algorithms with the intersection related operations,
and generates C++ implementations embedded with stream ISA assembly instructions.
The main challenge for code generation
is stream management (similar to register allocation in traditional compilers). 
Section~\ref{compiler} will discuss our solution in detail.

\rev{{\bf Applicability} The \projectname ISA is {\em flexible}---sufficient 
to support advanced optimizations used in
the leading GPM systems such as
Automine~\cite{mawhirter2019automine}, Peregrine~\cite{jamshidi2020peregrine}, and 
GraphPi~\cite{shi2020graphpi}.
We discuss these optimizations in 
Figure~\ref{fig:triangle_SB} and show 
how \projectname implements them in 
Figure~\ref{code:code_example_mining}.
Our ISA is also {\em general}: it not only accelerate\red{s} 
pattern enumeration but also 
the general tensor computation on sparse data.
It cannot benefit Pangolin~\cite{chen2019pangolin}, which 
is based on exhaustive check but the 
novel low-level APIs 
allow efficient filtering of infeasible
subgraphs and many optimizations. 
However, the users of Pangolin is expected to 
write GPM algorithms and optimizations using the 
APIs, thus the system is much more difficult to 
use. 
We strongly believe the compiler-based
approach has major advantage by hiding
all implementations details from the users. 
Moreover, it has been shown
that pattern enumeration method can also 
run faster than Pangolin on CPU (Table 5 of ~\cite{chen2020dwarvesgraph}). 
Finally, we emphaize that 
the code patterns of pattern enumeration 
\red{are not suited} for GPU.
In Section~\ref{perf_gpu},
we compare the performance of \projectname
with several manually
implemented algorithms in GPU.  }

We implement \projectname ISA and \red{its}
architectural components on zSim \cite{Sanchez-zsim}.
We use \red{seven} popular GPM applications
(triangle/three-chain/tailed-traingle counting, 3-motif mining, 4/5-clique counting, and FSM) on \red{ten} real
graphs.
\red{
The results show that \projectname significantly outperforms the \blue{InHouseAutomine} on CPU, 
on average \AvgSpeedupCPU and up to 
\MaxSpeedupCPU; and GRAMER, a state-of-the-art graph mining accelerator, based
on exhaustive check, on average \AvgSpeedupGRMER and up to 
\MaxSpeedupGRMER.
}

\section{Background}
\label{sec:Background}
\subsection{GPM Methods and Optimizations}
\label{sec:mining_basic}

A graph $G$ is represented by its vertex set $V$ and edge set $E$.
GPM applications take an input graph
and a pattern graph as inputs, enumerate all subgraphs matching a user-provided pattern, and extract useful information from them.
The subgraphs isomorphic to the pattern are named \textit{embeddings}.

\begin{figure}[h]
    \centering
    \vspace{-2mm}
\includegraphics[width=\columnwidth]{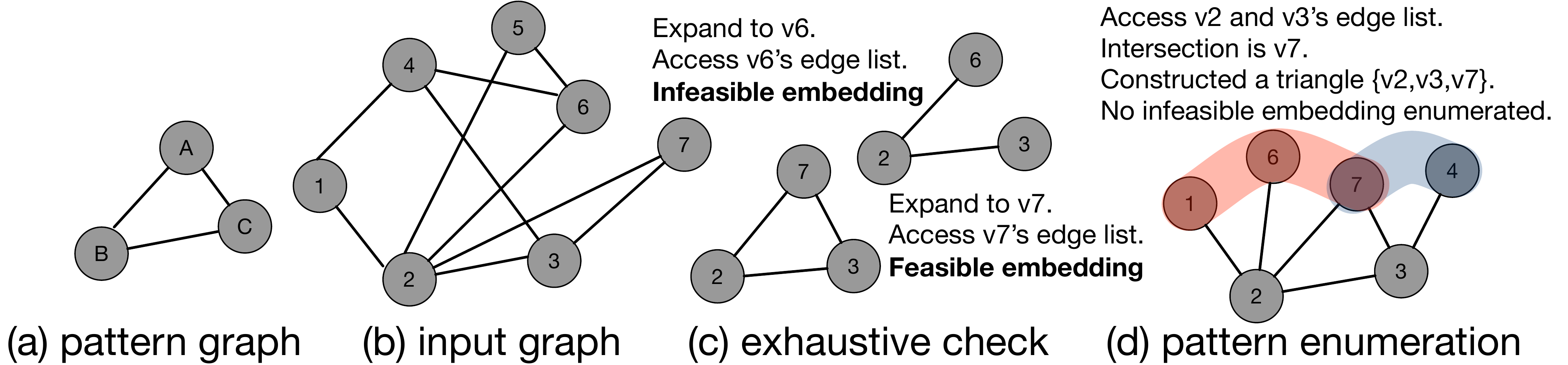}
    %
    \caption{\rev{Graph mining and two methods}}
    \vspace{-2mm}
    \label{fig:check_vs_enu}
\end{figure}
 

Figure~\ref{fig:check_vs_enu} shows the GPM problem which finds the pattern (triangle) (a)
in the input graph (b).
Figure~\ref{fig:check_vs_enu} (c) highlights
the subgraph expansion in exhaustive check. 
Suppose the current subgraph size is 2,
they are expanded into size-3 subgraph
from one of the vertices.
Then the new vertex's edge list is accessed
to construct size-3 subgraphs, some of them
are infeasible and can be filtered with 
user-defined filter function. 
It leads to excessive edge list 
accesses. 
Figure~\ref{fig:check_vs_enu} (d) shows
the memory access and computation of 
pattern enumeration. 
From two connected vertices, their edge lists
are accessed, followed by the {\em intersection} between
them. 
In this example, each common neighbor forms
a triangle embedding matching the pattern. 
While edge list accesses are still required, 
they are followed by the computation (intersection)
that is much more complex than graph computation
and cannot be efficiently performed in 
current processors. 
Moreover, since no infeasible subgraphs are
enumerated, the total number of edge list
accesses is also less.


\begin{figure}[!ht]
    \centering
    \vspace{-0mm}
    \includegraphics[width=\linewidth]{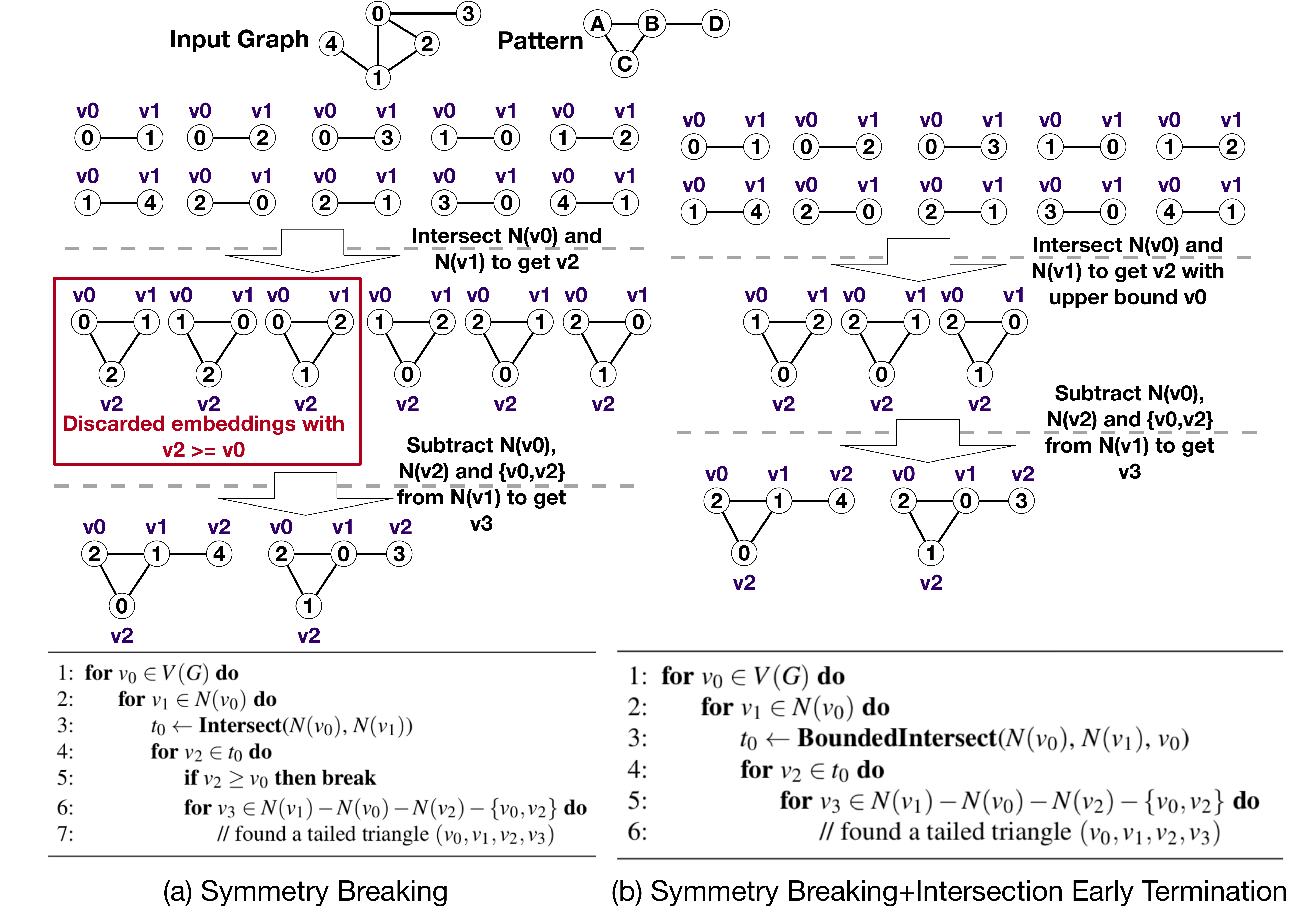}
    \vspace{-3mm}
    \caption{{Tailed-triangle mining}}
      \vspace{-5mm}
	\label{fig:triangle_SB}
\end{figure}

{
\textit{Symmetry breaking} in pattern enumeration avoids counting the same embedding for multiple times due to symmetry by enforcing a set of restrictions among vertices during embedding construction.
A tailed-triangle mining example is shown in Figure~\ref{fig:triangle_SB}.
We denote the first/second/third/forth matched vertex of an embedding as $v_0$-$v_3$.
Symmetry breaking requires $v_2<v_0$ 
so that the a unique embedding is enumerated 
only once, i.e., $(v_0,v_1,v_2,v_3)=(2,1,0,4)$ is the same as $(0,1,2,4)$.
As shown in (a), symmetry breaking first obtains all $v_2$ that is a common neighbor of $v_0$ and $v_1$ by intersecting $N(v_0)$ and $N(v_1)$, where 
$N(v)$ is the neighbor vertex set of $v$.
Then, it discards all $v_2$ that are no less than $v_0$ to satisfy the restriction (line 5-6 of the algorithm (a)).
This can be improved by \textit{early termination} of intersections since only the elements smaller than $v_0$ in $N(v_0)\cap N(v_1)$ are needed,
indicated as \texttt{BoundedIntersect()} in (b).
This optimization not only reduces the computation and 
accessed data in the edge list, but also
eliminate branches in the next loop level. 
}

\newcommand\tab[1][0.01cm]{\hspace*{#1}}

\subsection{Architecture Challenges}
\label{sec:arch_problem}

\begin{figure}
    \centering
    \vspace{-0mm}
    \includegraphics[width=\linewidth]{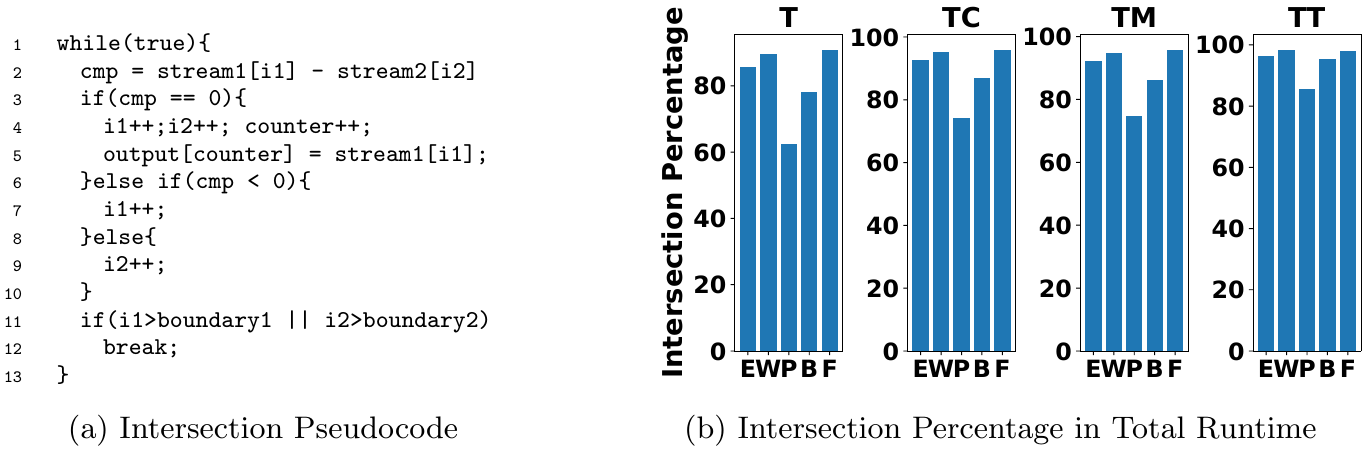}
     \vspace{-4mm}
    \caption{Intersection operations}
     \vspace{-6mm}
    \label{fig:breakdown}
\end{figure}

Figure~\ref{fig:breakdown} shows the 
example code that performs intersection 
operation. 
We abstract the edge lists as \texttt{stream1}
and \texttt{stream2}. 
If the end of streams have not been reached, 
the processor reads from \texttt{stream1}
and \texttt{stream2}, and compare the values. 
If they match, 
the processor advances the output pointer, 
writes back to the output array, 
advances pointers of \texttt{stream1} and 
\texttt{stream2}, and 
checks for boundary for both pointers. 
If the values  
mismatch, the processor advances 
the pointer of one of the streams, checks the boundary, fetches stream data, 
and compares again.
This code pattern contains branches and 
data dependencies in a tight loop,
making it difficult to predict the branches
and exploit instruction level parallelism. 
Figure~\ref{fig:breakdown} shows the 
percentage of cycles in a real machine for executing 
intersection operations for mining four patterns---Triangle(T), Three chain(TC), Three motif(TM), and Tailed triangle(TT)---on five data sets: patent(P), soc-sign-bitcoinalpha(B), email-eu-core(E), socfb-Haverford76(F), and wiki-vote(W). 
We see that intersection operations 
constitute major percentages of execution time.
These results justify our focus on 
improving the computation efficiency of 
GPM algorithms. 

\subsection{Can Existing Architectures Help?}
\label{sec:intersect_exist}

The support for intersection has
been proposed in the recent accelerator
architectures for DNNs and sparse tensors
because it is a key primitive to identify the 
effectual computations, e.g., the multiplication
of {\em two non-zeros}.
Specifically, Extensor~\cite{hegde2019extensor} 
uses intersection 
as the building block to construct sparse tensor algebra kernels, and 
developes a general architecture to efficiently
speed up tensor operation. 
SparTen~\cite{gondimalla2019Sparten} is a more specific sparsity-aware DNN accelerator that
performs dot-products of two vectors using intersection.
The two architectures differ in
(1) how the intersection operation is 
implemented: content addressable memory (CAM) based scan and search in 
Extensor versus prefix-sum in SparTen; and
(2) the generality: an architecture for the general sparse tensor computation (Extensor) or 
specifically targeting DNN acceleration (Sparten). 
On the other side, Sparse Processing Unit (SPU)~\cite{dadu2019towards,wang2019stream}
proposes specialized supports for stream-join (similar to intersection)
based on a systolic 
decomposable granularity reconfigurable architecture (DGRA). It uses pipeline to hide the latency of stream-join with 
a novel design of dataflow control model~\cite{dadu2019towards}.

However, none of the three architectures can 
efficiently support pattern enumeration.
First, the algorithm \rev{cannot 
be easily expressed in matrix operations.
The matrix-based GPM algorithm is
either for simple pattern~\cite{mattson2013standards} or 
for specific pattern~\cite{gleyzer2020leveraging}, 
and are not used in the state-of-the-art
general-purpose GPM systems. }
For this reason, it is almost impossible to port 
\rev{pattern enumeration} algorithm 
to Extensor or SparTen
which are specialized for tensor kernels 
and BLAS routines. 
Second, pattern enumeration relies on various
optimizations to achieve good performance, 
including computation reuse, 
symmetry breaking, and 
early intersection termination~\cite{mawhirter2019automine,mawhirter2019graphzero,shi2020graphpi,jamshidi2020peregrine}.
As shown in Figure~\ref{fig:triangle_SB},
they lead to complex code patterns 
that are infeasible to execute
efficiently without significant 
efforts on the more ``general'' specialized 
architectures such as SPU~\cite{dadu2019towards,wang2019stream}.
Specifically, SPU requires manually 
rewriting C codes
and describing data flow graph (DFG) with the language extensions for DGRA.
The computations are mapped to the systolic
DGRA by analyzing the DFGs. 
The complexity of pattern enumeration
algorithms leads to large DFGs, making 
it extremely difficult to port to SPU. 

GRAMER~\cite{yao2020locality} is the 
first GPM architecture.
However, it is designed for the 
much slower exhaustive-check method.
The key idea is using a heuristic algorithm 
to classify 
graph data into high and low priority by 
approximately simulating the vertex and edge
accesses during exhaustive-check in a pre-processing step. 
Then the high-priority data are permanently
stored in the fast memory while the low-priority
data are organized in the cache with a 
specialized data replacement policy.
GRAMER achieves impressive speedups
compared to two recent GPM
frameworks on CPU---$1.11\times \sim 129.95\times$ over RStream~\cite{wang2018rstream} and Fractal~\cite{dias2019fractal}.
Due to the large performance gap,
GRAMER running exhaustive check method 
is very likely to be slower than 
pattern enumeration on unmodified CPU. 
Even for small patterns like triangle counting, Automine can be $68.6\times$ faster than RStream; for more complex patterns, it can achieve $777\times$ speedup. 
This comparison reveals the importance of
the superior algorithm, and it is critical
to develop architectural supports based on 
the state-of-the-art method.

\subsection{Design Principle}
\label{sec:mot}

Our design principle is to develop architectural
supports based on the conventional processor, 
instead of designing an accelerator for 
graph mining from the ground up. 
It is justified by the complex control 
flows and code patterns of the state-of-the-art
pattern enumeration algorithms with advanced optimizations.
In the next two sections, 
we will describe our novel 
{\em vertical approach, \projectname}, from the
instruction set extension for streams (Section~\ref{isa_extension}) to the architecture components 
that implement the new instructions (Section~\ref{sec:arch}).

\section{Stream ISA}
\label{isa_extension}

For GPM, the key operation is the 
intersection between two sparse vectors, 
e.g., edge lists.
In general, we define a sparse vector 
as a {\em stream},
which can be:
(1) a {\em key stream}---a list of {\em keys}, such as the edge
list in graph representation; or 
(2) a {\em (key,value) stream}---a list of {\em (key,value)}, such as the pair
of indices of non-zero elements and
their values in a sparse matrix representation. 
We propose a novel instruction
set extension that {\em intrinsically operates
on streams}, supporting both data movement and computation. 
The proposed stream ISA extension can be 
considered as a natural extension to 
the traditional instructions 
for ordinary scalar values.

\subsection{Register Extension}

The stream ISA extension represents stream
as the first-class data type. 
The processor uses $N$
{\em stream registers} to maintain stream 
information, where $N$ is the maximum number
of active streams supported.
A stream is {\em active} between its initialization
and free---each can be performed by 
an instruction.
A stream register stores 
the stream ID, the stream length, the start key
address, the start value address, 
and a valid bit. 
The stream registers cannot be 
accessed by any instruction 
and is setup up when the corresponding stream is initialized. 
The program can refer to a stream by 
the stream ID, the mapping
between a stream ID and its stream register
is managed internally in the processor with
the {\em Stream Mapping Table (SMT)}
(see details in Section~\ref{sec:sreg}).
The key and value address of a stream register
are only used
by the processor to refer to the keys and values
when the corresponding stream ID is referenced.


We also add three registers to keep the 
information about compressed sparse row (CSR) graph format {\cite{bulucc2009parallel}}. 
They hold pointers to CSR index, CSR edge list, and CSR offset and can be initialized by 
an instruction.
The CSR offset stores the offset of the 
the smallest element larger than the vertex itself in the neighbor list. 
It is used to support the nested intersection, and the symmetric breaking optimization. 
The design can be adapted to other sparse 
representations. 


\subsection{Instruction Set Specification}
\label{sec:isa}

\begin{table*}[t]
\resizebox{2.08\columnwidth}{!}{%
	\begin{tabular}[t]{l|l|l}
	\hline
	Instruction  & Description & Operands   \\
	\hline
	\texttt{S\_READ	R0, R1, R2} & Initialize a key stream & R0:start key address, R1:stream length, R2:stream ID\\\hline
	\texttt{S\_VREAD R0, R1, R2, R3} & Initialize a (key,value) stream & R0:start key address, R1:stream length, R2:stream ID, R3:start value address\\\hline
	\texttt{S\_FREE R0}	& De-allocate a stream &  R0:stream ID             			   \\\hline
	\texttt{S\_FETCH R0, R1, R2}		& Return one element of a key stream &  R0:stream ID, R1:element offset, R2: returned element       		\\\hline
	\texttt{S\_SUB	R0, R1, R2, R3} 	& Subtraction of two streams, use stream of id R0 to subtract stream of id R1  & R0,R1: input stream IDs, R2:output stream ID, { R3: upper-bound of the subtracted result} \\\hline
	\texttt{S\_SUB.C	R0, R1, R2, R3} 	&  Return \# of elements in subtraction of two streams, use stream of id R0 to subtract stream of id R1 & R0,R1: input stream IDs, R2:returned result, {R3: upper-bound of the subtracted result}  \\\hline	
	\texttt{S\_INTER	R0, R1, R2, R3} & Intersection of two streams &  R0,R1: input stream IDs, R2:output stream ID, {R3: upper-bound of the intersected result}\\   \hline
	\texttt{S\_INTER.C	R0, R1, R2, R3} & Return \# of elements in intersection of two streams&  R0,R1: input stream IDs, R2: returned result, {R3: upper-bound of the intersected result} \\   \hline	
	\texttt{S\_VINTER	R0, R1, R2, IMM} & Sparse computation using the values of two (key,value) streams &  R0,R1: input stream IDs, R2:returned result, IMM: specify user-defined op \\  \hline 
    \texttt{S\_CSR R0, R1, R2} & Register pointers to CSR graph structure & R0:CSR index address, R1:CSR edge list, R2: CSR offset \\     \hline
	\texttt{S\_NESTINTER R0, R1} & Nested intersection & R0: stream ID, R1: returned result\\         
	\hline
	\end{tabular}
	}
	\caption{Stream ISA Extension. R0-R3 are general-purpose registers, IMM is an immediate value.}
	\vspace{-4mm}
	\label{table:instructions}
	
\end{table*}

Table \ref{table:instructions} lists
the instruction set extension for 
streams. 
The instructions can be classified
into three categories:
(1) stream initialization and free;
(2) stream computation; and 
(3) stream element access. 
The input operands for all 
instructions are general purpose registers
containing stream ID. 
There is no reason immediate values
cannot be used directly as the inputs---we just assume the register operands for 
simplicity, the same architecture can support both scenarios.

\texttt{S\_READ} and \texttt{S\_VREAD}
are the instructions to {\em initialize}
a key stream and (key,value) stream, 
respectively. 
The operands are general purpose 
registers containing start key address 
(also start value address for \texttt{S\_VREAD}), stream length, and
stream ID.
After they are executed, if the stream ID 
is not active, an unused stream register (valid bit is 0)
will be allocated to the stream and the 
new mapping entry is created and inserted
into SMT. 
If the stream ID is already active, 
the previous mapping is overwritten with 
the current stream information. 
After creating the mapping to a stream 
register, both instruction will also 
trigger the fetching of {\em key stream}
to the stream cache (see details in Section~\ref{sec:scache}).
Thus, if the current stream overwrites
the previous one, the content in the 
stream cache will also be updated. 
Note that \texttt{S\_VREAD} does not
load the values, which will be
triggered when the computation 
instruction for (key,value) stream (\texttt{V\_VINTER})
is executed. The values are accessed and 
fetched through the ordinary memory hierarchy rather than the stream cache.
\texttt{S\_FREE} is used to free a stream. 
When it is executed, the processor 
finds the SMT entry for the stream ID 
indicated in the operand and set the 
valid bit to 0. 
If such entry is not found, an exception
is raised.




Our ISA extension contains six instructions
for {\em stream computation}.
\texttt{S\_INTER}, 
\texttt{S\_INTER.C},
\texttt{S\_SUB}, 
\texttt{S\_SUB.C} perform 
the simple computation 
on key stream---intersection and subtraction.
The suffix ``\texttt{.C}'' indicates the
variants of the corresponding 
instructions that do not 
output the result stream but just the 
{\em count} of non-zeros in the result
stream. 
\blue{If the output is a stream, the stream ID of an initialized stream should be given in one of the input registers. The stream ID will then be added into SMT.}

\rev{All these instructions take a upper-bound operand \texttt{R3} to support early intersection/subtraction termination.
Once all output stream elements smaller than \texttt{R3} have been produced, the instruction terminates the computation early.
For unbounded operations, 
\texttt{R3} is set to $-1$. It is used to 
implement the early terminate optimization shown in Figure~\ref{fig:triangle_SB} (b). The conditional 
termination can be easily implemented inside
the intersection unit. 
}
Next, we explain the two  
complex instructions. 

The first one is \texttt{S\_VINTER}, which 
performs the user-defined intersected value
computations. 
The instruction first computes
the {\em intersection of the keys} of the two 
input (key,value) streams, and then
performs the {\em computation on the values}
corresponding keys. 
For example, the key intersection of 
two (key,value) streams
$[(1,45),(3,21),(7,13)]$ and \\
$[(2,14),(5,36),(7,2)]$ is 7.
The instruction performs the computation on the 
corresponding values: assuming the computation
is multiply-accumulation (\texttt{MAC}) specified in \texttt{IMM},
the result is 13 $\times$ 2 = 26 in \texttt{R2}.
The other types of computation can be specified
in \texttt{IMM}, such as \texttt{MAX} (choose the maximum and accumulate), \texttt{MIN} (choose the minimum and accumulate), or any reduction operation. 
In \projectname architecture, the computation 
on values is performed by a dedicated functional 
unit, which can be easily extended to perform
new operations. 
This instruction is useful in sparse matrix
computation, where the keys indicate the 
positions of the non-zeros and the 
actual computations are performed on these values.
It is not used in GPM since there is 
no value involved. 
If any input stream ID is 
not a (key,value) stream, an exception is
raised. 

The second complex instruction is \texttt{S\_NESTINTER}, which performs the 
{\em nested intersection}.
It is an instruction specialized for GPM.
Let the input stream (an edge list) be $S=[s_0,s_1,...,s_k]$, where each $s_i$ corresponds to a vertex. 
Let us denote the edge list of each $s_i$ as
$S(s_i)$, and the result of the instruction as
$C$.
This instruction performs the following
computation: $C=\sum_{i=0}^{i=k}count((S \cap S(s_i)))$,
where $\cap$ is the intersection between 
two key streams, {and $count$ returns the 
length of a stream.} \blue{The intersections are bounded by the value of $s_i$.}
Thus, this instruction implements
a kind of {\em dependent stream intersection}.
Given a stream $S$, the other streams
to be intersected with it are determined 
by the keys (vertices) of $S$.
The generation of the dependent streams
corresponding to each $s_i$ is performed 
by the processor using the information
in the three CSR registers, which are 
loaded once using \texttt{S\_CSR} before processing a graph.

The \texttt{S\_FETCH} instruction performs the 
{\em stream element access}---returning
the element with a specific offset in a stream,
which can be either the output stream of an intersection operation 
or \blue{ an initialized stream loaded 
from memory. }
Typically, the offset is incremented
to traverse all elements in a stream. 
When it reaches the end of the stream, 
\texttt{S\_FETCH} will return a special
``End Of Stream (EOS)'' value. 

\subsection{Code Examples}


\begin{figure}[t]
    \centering
    \vspace{-0.1cm}
\includegraphics[width=\columnwidth]{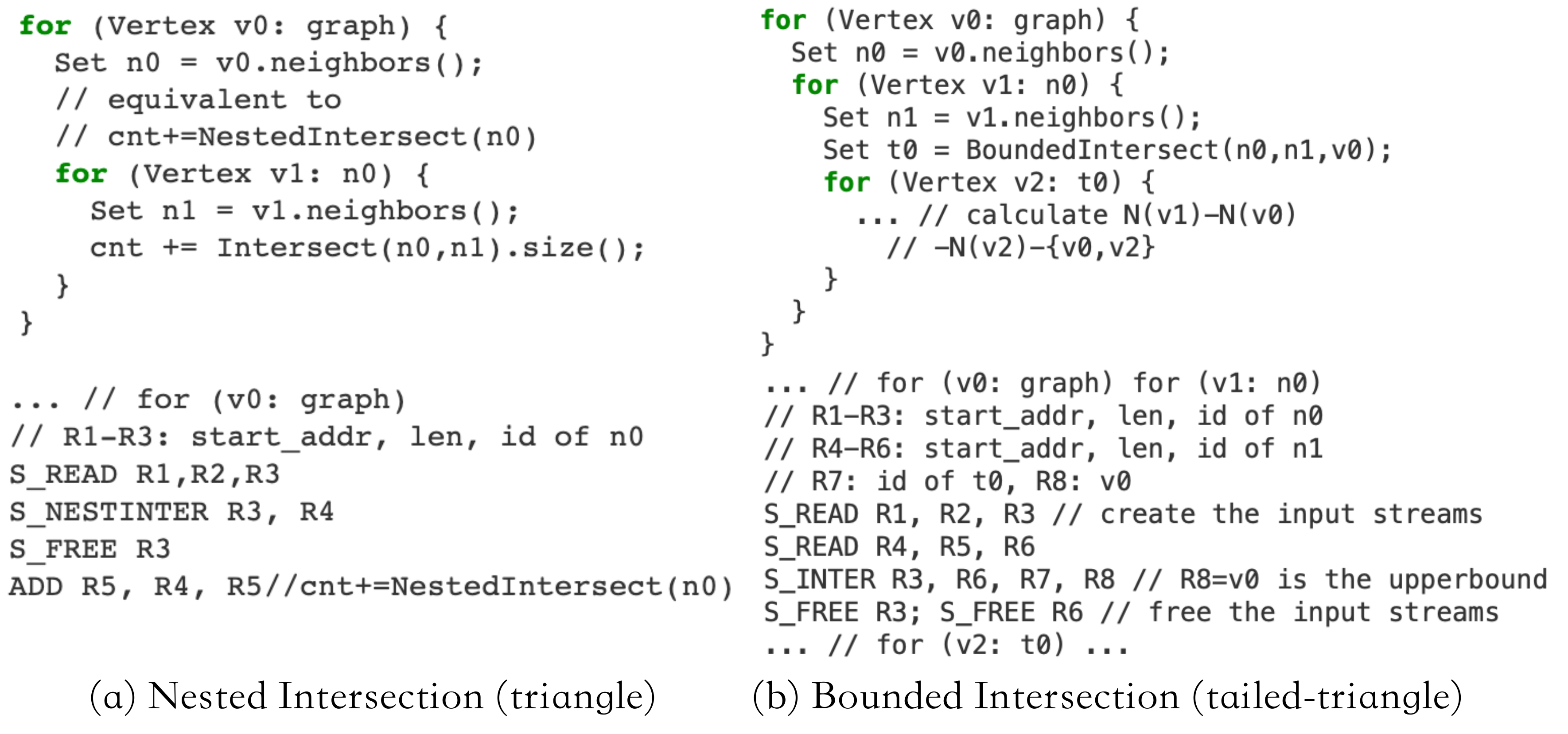}
        \vspace{-3mm}
        \caption{\rev{Pattern enumeration with Stream ISA Extension}}
        \vspace{-4mm}
        \label{code:code_example_mining}
\end{figure}


Figure~\ref{code:code_example_mining} (a) shows how to implement triangle counting using our ISA extension. 
The \texttt{v1} for-loop is essentially a nested intersection operation on \texttt{n0}.
Thus, we can nicely 
use \texttt{S\_NESTINTER} to implement it.
There is only one active stream
whose ID is stored in \texttt{R3}.
The multiple intersections performed by \texttt{S\_NESTINTER} do not take 
stream \blue{register} resource. 
In addition, there is only one level of
loop in the assemble code. 
Such specialization based on the 
understanding of the GPM code 
pattern is critical to achieving high
performance. 
\rev{
Figure~\ref{code:code_example_mining} (b) shows the implementation of tailed-triangle mining (shown in Figure~\ref{fig:triangle_SB} (b)) with intersection early termination.
The intersection inputs \texttt{n0} and \texttt{n1} are loaded into two streams with IDs \texttt{R3} and \texttt{R6}, respectively.
We then use \texttt{S\_INTER} to intersect them with an upper-bound \texttt{v0} (stored in \texttt{R8}) so that the intersection can terminate early.
}

Note that our ISA allows different loop iterations to 
use the same stream IDs, similar to the 
same variable names. 
The processor internally keeps track
of the active streams in both front-end (after instruction decoding) and back-end
(at instruction commit time), and 
will recognize the same stream IDs in 
different iterations as different streams.

\begin{figure}[ht]
     \vspace{-0.3cm}
    \centering
\includegraphics[width=\columnwidth]{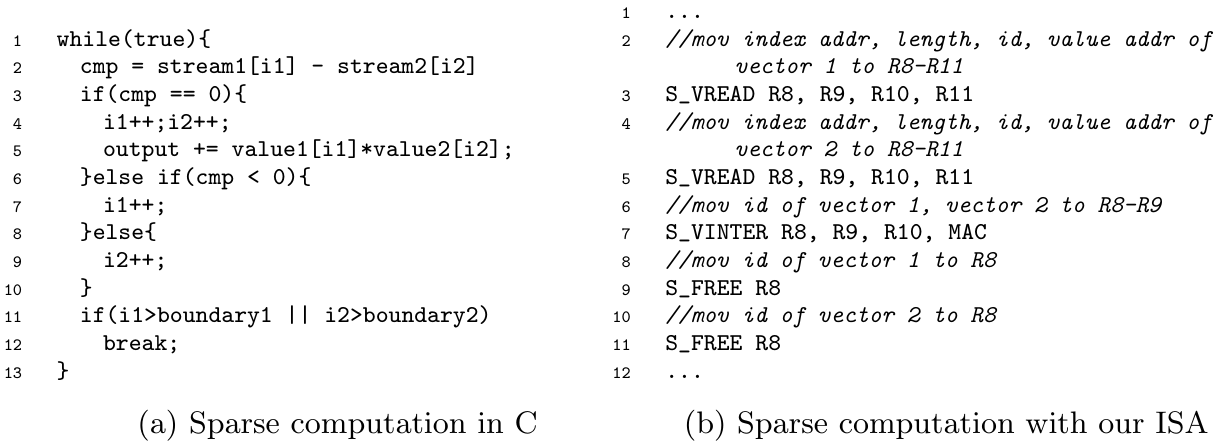}
        \vspace{-0.3cm}
        \caption{Sparse vector multiplication with Stream ISA Extension}
         \vspace{-0.4cm}
        \label{code:svm_example}
\end{figure}

Figure \ref{code:svm_example} (a) and (b)
show the vector multiplication
implementation with our stream ISA extension. 
At line 3 and 5, two
(key,value) streams are initialized
using the \blue{addresses} of
two sparse vectors with \texttt{S\_VREAD}.
Line 7 performs the multiply-accumulation
on the values of the intersected keys.

\section{\projectname Architecture}
\label{sec:arch}

\subsection{Overall Architecture}
\label{sec:proc}

\begin{figure}[tb]
\vspace{-0mm}
\centering
\includegraphics[width=\columnwidth]{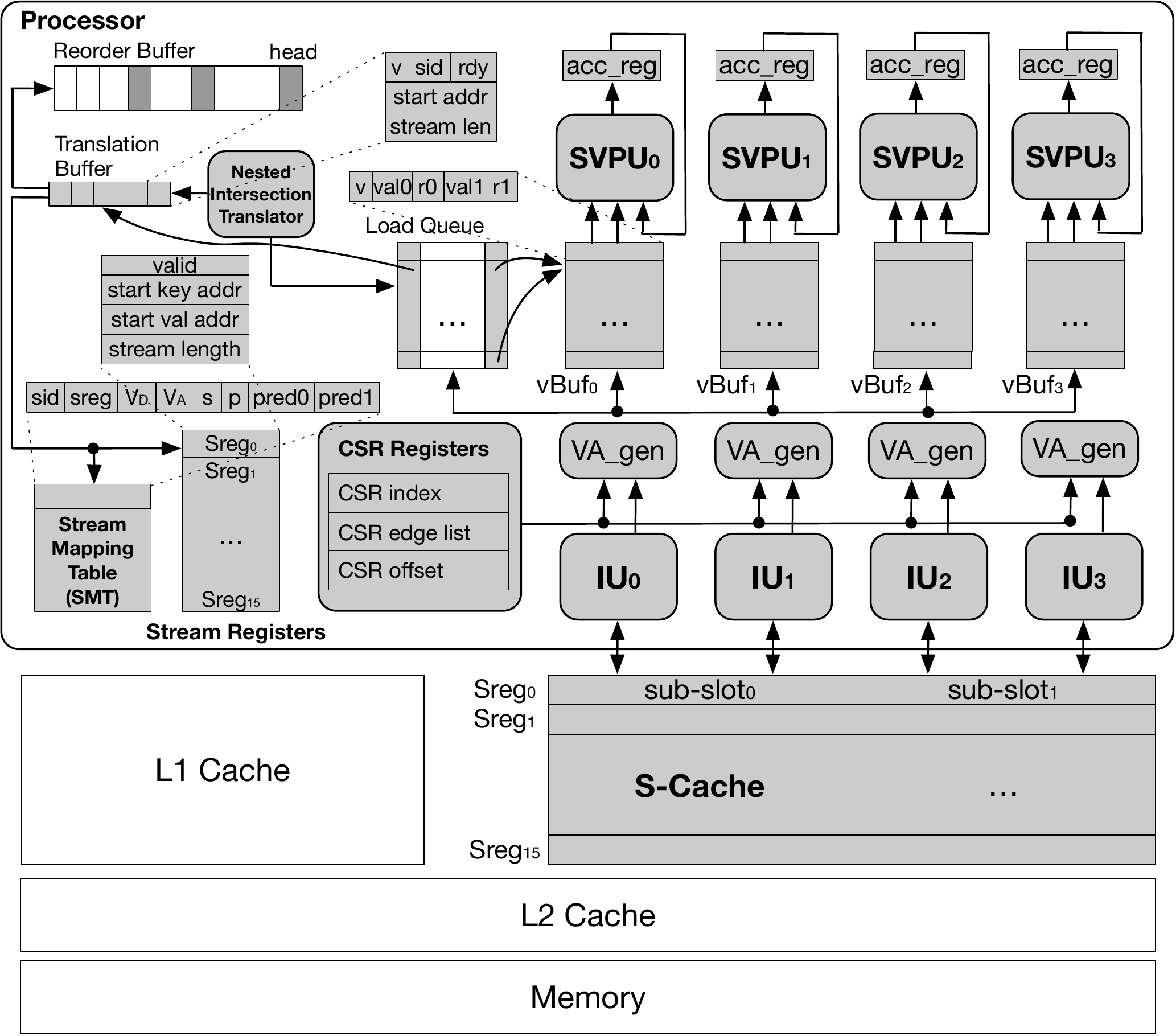}
\vspace{-2mm}
\caption{\projectname Architecture}
\vspace{-10mm}
\label{fig:overall}
\end{figure}

The \projectname architecture is composed of specialized structures
built on conventional processor architecture and memory hierarchy that 
implement the stream ISA extensions. 
Figure~\ref{fig:overall} shows a detailed overview with 
stream related components highlighted in gray color. 
All instructions in Table~\ref{table:instructions} except \texttt{S\_NESTINTER}
occupy one entry in the Reorder Buffer (ROB). 
To support the ISA extension, the architecture needs to 
solve a number of problems:
(1) the mapping between stream ID and stream register, which is handled 
by the Stream Mapping Table (SMT);
(2) the movement of stream data, which is supported efficiently by 
the stream cache (S-Cache);
(3) the dependency between streams, which is tracked
with a property of intersection
and minor supports in S-Cache;
(4) the implementation of \texttt{S\_VINTER}, which is realized
by the coordination among the Intersection Unit (IU),
Stream Value Processing Unit (SVPU), and the load
queue augmented with stream information;
(5) the implementation of \texttt{S\_NESTINTER}, which is realized 
by the nested intersection translator that generates the micro-op sequence, similar
to the contemporary implementation of CISC instructions 
with RISC-style micro-ops.






\subsection{Stream ID Mapping}
\label{sec:sreg}


In \projectname, each stream ID ({\em Sid}) specified in an instruction is mapped
to an internal stream \blue{register} ({\em Sreg}). 
This mapping is performed at the front-end after instruction decoding 
and the mapping relation is kept in SMT.
Besides the stream ID and its mapped stream 
register, each SMT entry contains:
(1) two valid bits: $V_D$, indicating the {\em define} point of the stream, 
and $V_A$, indicating whether the stream is {\em active}; 
(2) the {\em start (s)} and {\em produced (p)} bit, which 
indicate whether S-Cache contains the keys
from the start of the stream and whether the data
for the whole stream is produced (so that it can be 
used by the dependent streams); and
(3) the {\em pred0} and {\em pred1}: the 
IDs of the \blue{streams} that the current stream depends
on.
In this section, we explain the two valid bits
and the others will be discussed together 
with S-Cache and dependence handling. 

Initially, both $V_D$ and $V_A$ are 0 and SMT is empty.
Both $V_D$ and $V_A$ are 
set after decoding a \texttt{S\_READ} or a \texttt{S\_VREAD} 
instruction and the SMT entry indicates that the \texttt{Sid\_i} 
in the last operand of the instruction is mapped to \texttt{Sreg\_j}.
Both $V_D$ and $V_A$ are set to one, they indicate that 
the instruction defines $Sid_i$ and it is active. 
Later, when \texttt{S\_FREE Sid\_i} is decoded, 
the SMT is examined and an entry for \texttt{Sid\_i} should be found (otherwise
an exception is raised), and its $V_D$ is reset, while $V_A$ is unchanged. 
This means that \texttt{Sid\_i} is no longer defined---the instructions
after \texttt{S\_FREE Sid\_i} should not be able to reference \texttt{Sid\_i}---but
the stream is still active since \texttt{S\_FREE Sid\_i} has not
been retired.
When \texttt{S\_FREE Sid\_i} is retired, $V_A$ is reset and 
the entry becomes free. 
When a new stream is mapped, 
the processor checks SMT and finds an entry with $V_A=0$, which implies $V_D=0$. Note that is not true vise versa---$V_D=0$ does not
imply $V_A=0$.

Our design expects the codes to call \texttt{S\_FREE} after a 
stream is no longer used, so that its SMT entry can be released. 
In Section~\ref{sec:api} we will describe the APIs for programmers, who do not need to
directly write assembly codes, thus this requirement can be easily achieved. 
When all stream registers are occupied ($V_A=1$), the instruction that initializes
a new stream will be stalled. The larger (or even unlimited)
number of stream IDs can be 
supported by virtualization---by saving some SMT entry to a special memory
region to release SMT space. Due to the space limit, we do not discuss
this in detail. In fact, using 16 stream registers is enough for 
all our applications.
The design can naturally support the stream operations in loop 
iterations. 
Typically, inside an iteration, some streams are initialized and 
computations on them are performed before \texttt{S\_FREE}s at the end of 
the iteration (refer to Figure~\ref{code:code_example_mining} (c) for an example).
The different iterations can use the same stream IDs, which will be mapped to 
different SMT entries with our SMT mechanisms. 

Note that the SMT mechanisms
will not increase the latency of 
CPU pipeline, and they can be 
implemented in a pipelined manner
similar to the register rename stage
in CPU. 
Specifically, the mapping from 
architecture registers to physical registers is similar to the mapping
from {\em Sids} to {\em Sregs}. 
We also maintain the ``readiness'' of
stream IDs.

\subsection{Stream Cache} 
\label{sec:scache}

In \projectname, the keys for each active stream are loaded into a special 
{\em stream cache (S-Cache)}, which is on top of L2 cache together with L1. 
Note that the values in (key,value) stream 
are still fetched through the 
normal memory hierarchy.
Thus, when the stream keys are accessed using the stream instructions, 
the data will not pollute L1. 
Such specialization enables efficient stream 
data movements while avoiding cache evictions by other data. 
Since the keys of a stream are accessed sequentially, the data can be 
effectively prefetched to S-Cache without a complex prefetcher, thanks to 
the {\em known} access pattern.
The organization of S-Cache is simple: each stream register has 
a {\em slot} that holds a fixed number 
keys of the stream. 
We use the 64-key slot which leads to 256 byte slot size. 
With 16 stream registers, the total size of S-Cache is 
4KB. 

When an \texttt{S\_READ} is executed, 
the first 64 keys are fetched to the S-Cache,
and the {\em start} bit in SMT for the 
stream is set. 
Unless the length of the stream is \blue{no} 
more than 64, at this point the S-Cache only contains the first portion of the stream. 
The {\em start} bit indicates that the instructions
that depend on the stream can use the data in the 
S-Cache slot. 
Referring to Table~\ref{table:instructions}, 
our ISA does not contain any instruction
that explicitly stores to a stream:
only \texttt{S\_INTER} and \texttt{S\_SUB}
produce the results in the destination stream. 
When these instructions are executed, 
the result keys are written to the S-Cache
slot in group of 64. 
If the result stream contains more than 64 keys,
the slot will contain {\em the most recently 
produced 64 keys} while the previous slot
is written back to L2 and the {\em start} bit is cleared. 
When the whole result stream is generated 
by the computation instruction, the {\em produced}
bit is set, which is used to trigger the dependent 
instructions. 

The typical code pattern is that two streams
are initialized by \texttt{S\_READ} before the intersection operation is performed.
In this case, data fetching 
from L2 to S-Cache and transfer to IUs for computation can be pipelined. 
To support that, we use the idea of 
double buffer and divide each slot into two sub-slots. 
When a sub-slot is fetched from L2, the keys in the other sub-slot
can be prefetched to IU simultaneously and the intersection computation
can be overlapped.
We assume that the bandwidth between the stream cache and IU is 4 keys (16 bytes)
per cycle, which is similar to the read bandwidth of L1 cache of 
Intel Nehalem at 128 bits per cycle~\cite{intel}.

Typically, the intersection computation time of a sub-slot 
(e.g., 32 to 64 cycles) is longer than fetching data from L2
on cache hits (e.g., 20 cycles). 
With multiple IUs, the parallel execution time of 
multiple intersections can be better overlapped
with the data fetching time of
these streams.
When multiple IUs (4 in our design) need data to 
perform computations, S-Cache has to schedule the 
data transfer to different IUs. 
We use a simple round-robin policy:
at each cycle, S-Cache schedules the transfer of 4 keys 
to a different IU that is waiting for the data.
Each IU is able to perform the intersection on the partial 
key streams received.

\subsection{Stream Data Dependency}

Two streams may have dependency due to:
(1) stream ID, where an instruction uses the output
stream of a previous computation 
(\texttt{S\_INTER} or \texttt{S\_SUB}) \blue{as an input stream}; or
(2) the overlapped memory regions of two streams. 
It is easy to handle the first scenario: 
after the stream IDs are available after
decoding, the dependency can be handled in the similar manner
to the data dependency on general registers. 
When a dependency is identified, 
the consumer instruction can only execute
after the producer instruction. 
It is enforced by filling the {\em pred0} and
{\em pred1} in SMT of the consumer instruction. 
When the producer instruction finishes,
its SMT entry's {\em produced} bit is set. 
Each cycle the processor checks the status
of the producer instruction(s) and triggers
the consumer instruction when all operands' {\em produced} bit 
are set. 
If the key stream produced is less than 64 keys, 
the whole stream is in S-Cache with the {\em start}
bit set, the consumer instruction reads
directly from S-Cache; otherwise, the slot
will be refilled from L2. 

For the second scenario, we can check the 
potential dependency conservatively by 
leveraging the fact that {\em the 
length of the output stream is 
less than the minimum length of the two input streams}. 
Thus, we can conservatively deduct
the maximum length of 
the output stream. 
The possibly overlapped stream memory regions
can be detected using the start key address and 
stream length of different streams.  
The dependent stream instructions need to be
executed sequentially, which is 
enforced using the same mechanism as the 
first scenario.

\subsection{Sparse Computation on Values}
\label{sp_value}

The sparse computation on values is suppored by the 
coordination between IU, value buffer (vBuf), load queue, and 
Stream Value Processing Unit (SVPU). 
When \texttt{S\_VINTER} is executed, 
an IU starts with key intersection calculation and the output
keys are given to the {\em Value Address Generator (VA\_gen)}
associated with the IU (refer to Figure~\ref{fig:overall}).
\red{VA\_gen generates the value addresses for each key in the intersection.} 
These addresses are sent to load queue to request the values 
through the normal memory hierarchy, rather than S-Cache. 
Each value request is also allocated with an entry 
in the vBuf, which will collect the two values returned 
from the load queue (val0 and val1). 
Each entry has a ready bit (r) for each value, which is set
when the load queue receives the value.
We assume that the operation is commutative (e.g., multiply-accumulate)
thus the 
computation using val0 and val1 can be performed by SVPU as soon as both 
ready bits are set.
We do not need to enforce any order \blue{on the accumulation}. 
The {\em acc\_reg} is used to keep the accumulated partial results. 
While performing substantial amount of computations, 
this instruction only takes one entry in ROB. 
\blue{After} the final result 
is produced in the acc\_reg of the corresponding IU, 
it will be copied to the destination register, and \blue{then} the instruction \blue{will} retire from the processor when it reaches the head of ROB.

\subsection{Nested Intersection}
\label{nested}

The \texttt{S\_NESTINTER} is the most complex instruction 
and we use the {\em Nested Instruction Translator} to generate
the instruction sequence of other instructions of stream ISA to implement it. 
Based on the input key stream, the translator first generates
the stream information based on each key element. 
\blue{The memory addresses of the streams information are calculated based on the CSR registers,}
then the memory requests are sent through load queue.
For each stream,
an entry is \blue{allocated}
in the {\em translation buffer}, its ready bit (rdy)
is set when the stream information 
is returned at load queue.
Similar to the pointer to vBuf entry, each load queue entry also keeps a pointer
to the translation buffer entry. 
For each nested stream, three instructions are generated:
\texttt{S\_READ}, \texttt{S\_INTER.C}, and \texttt{S\_FREE}.
An addition instruction is generated to accumulate the 
counts. 
Each instruction takes an entry in the translation buffer. 
The start address and stream length fields are only 
used in \texttt{S\_READ}. 
When the stream information is ready, the three instructions are inserted into ROB.
The translation is stalled when the translation buffer is full, which
can be due to either ROB full or waiting for the stream information. 
In either case, the space will later be released because eventually 
the instructions in ROB will retire and the requested data
will be refilled. 
These events do not wait for the translation procedure \blue{thus}
there is no deadlock.

\section{Implementation and Software \\ Interface}
\label{sec:impl}

\subsection{Implementation Considerations}

The \texttt{S\_NESTINTER} is translated into a variable length
instruction sequence by the 
\blue{{\em Nested Instruction Translator}}
and will take multiple 
ROB entries. To ensure the precise exception, the processor takes
a checkpoint of registers before the instruction. 
If an exception is raised during the execution of the instruction sequence,
the processor rolls back to the checkpoint and raises the exception handler.
It is similar to the mechanisms for transactional and atomic block 
execution~\cite{ceze2006bulk,qian2014omniorder}.
Besides the normal information such as general registers, 
the checkpoint includes the content of 
SMT, stream registers and CSR registers.
Another assumption in \projectname architecture is that
the stream cache does not participate in the coherence protocol.
Thus, the potential modifications of 
other cores on key elements do not propagate 
in time to stream cache. 
However, for the applications that \projectname 
\blue{is targeted for, the data (such as graph or sparse matrix) are read-only.}
Therefore, it \blue{does not cause any major problem.}

\rev{
\blue{In \projectname, due to the complex code patterns in pattern enumeration, we {\em tightly couple} the accelerator units with the processor, rather than building a stand-alone accelerator. }
A more decoupled architecture would incur
relatively high overhead for the interaction 
between CPU and the accelerator. 
Nevertheless, with more 
significant change of the software and 
an appropriate way to map each function to architecture,
it is possible to develop a more decoupled 
accelerator. We leave that as the future work.}

\subsection{\rev{Hardware Cost}}
\label{hw_cost}

\begin{figure}[h]
     \vspace{-0.6cm}
\begin{lstlisting}[basicstyle=\tiny,style=MyScala]
for(Vertex v1 : graph){
 Vertex* addr = &v1.neighbor();
 Length len = v1.neighbor().size();
 VertexSet set1 = RegisterVertexSet(addr, len);
 Length counter = 0;
 while(counter<len){
  Vertex v2 = EnumerateVertexSet(set1,counter);
  Length len2 = v2.neighbor().size();
  Vertex* addr2 = &v2.neighbor();
  VertexSet set2 = RegisterVertexSet(addr, len);
  patternNum += SubtractVertexSetCount(set1,set2);
 }
}
\end{lstlisting}
    \vspace{-0.2cm}
        \caption{API Example: Three Chain}
        \vspace{-0.4cm}
        \label{code:full_example}
\end{figure}

\rev{The Coordinator module from ExTensor~\cite{hegde2019extensor} and the IU of \projectname have similar functionality---performing intersection logic, and if anything the Coordinator should be more complex. Thus we substitute IU area with Coordinator area. ExTensor lists the total area of Coordinators to be $2.38mm^2$ using 32nm, and with 129 individual coordinators (128 from PE and 1 in LLB), each coordinator is thus 0.0184mm2. 
The cost of 4 IU in \projectname
is thus $0.0738mm^2$.
Similar to ExTensor, we use CACTI~\cite{cacti} 
to model area for its SRAM components, and stream registers, stream mapping table, and stream cache. At 32nm and with implementation as scratchpad RAM, stream register file takes $0.0008mm^2$, stream mapping table $0.0010mm^2$, and stream cache $0.0175mm^2$, for a total of $0.0193mm^2$.
In total, the most additional 
area for \projectname (memory and intersection logic) is around $0.0931mm^2$ at 32nm.
Of course, actual integration into the cores would require further routing which will increase this number.
However we show that the \projectname hardware modules do not constitute a significant hardware cost.}

\label{sec:api}
\begin{table}[t]
\resizebox{1.00\columnwidth}{!}{%
\begin{tabular}[t]{l}
	\hline
\textbf{VertexSet}  {RegisterVertexSet} (\textbf{Vertex*} addr, \textbf{Length} len )\\
\textbf{void }{ReleaseVertexSet}   (\textbf{VertexSet} handler)  \\
\textbf{Length }{NestCounting}   (\textbf{VertexSet} handler)  \\
\textbf{Vertex } {EnumerateVertexSet}   (\textbf{VertexSet} handler, \textbf{Length} offset)          		\\
 \textbf{void }{SubtractVertexSet}   (\textbf{VertexSet} A, \textbf{VertexSet} B, \textbf{VertexSet} C, \textbf{VertexId} D)  \\           
\textbf{Length }{SubtractVertexSetCount}   (\textbf{VertexSet} A, \textbf{VertexSet} B, \textbf{VertexId} C)  \\           
\textbf{void} {IntersectVertexSet}  (\textbf{VertexSet} A, \textbf{VertexSet} B, \textbf{VertexSet} C, \textbf{VertexId} D)    \\
\textbf{Length }{IntersectVertexSetCount}  (\textbf{VertexSet} A, \textbf{VertexSet} B, \textbf{VertexId} C)    \\
 \textbf{Vector} {RegisterVector} (\textbf{Index*} addr, \textbf{Value*} addrV, \textbf{Length} len)     \\
 \textbf{void }{ReleaseVector}  (\textbf{Vector} handler) \\
 \textbf{Value} {VectorCompute}  (\textbf{Vector} A, \textbf{Vector} B, \textbf{Op} type)  \\
	\hline
  \end{tabular}
  }
	\caption{\projectname APIs (Function Prototype)}
	\vspace{-7mm}
	\label{table:api}
\end{table}

\subsection{GPM Compiler and Low-Level APIs}
\label{compiler}

\projectname provides high-level software interface for the typical users of GPM systems, such
as scientists and data analysts, who are not
algorithm experts. 
The key is to provide good programmability.
To use the high-level interface, the users
just need to specify the patterns and the
input graph, and the compiler
will generate the binary for mining the given 
pattern(s). 
Similar to Automine~\cite{mawhirter2019automine},
we developed a GPM compiler 
to generate stream ISA based GPM implementation generation. 
The compiler takes the user-specified patterns as input, 
synthesizes the corresponding intersection based GPM algorithms (e.g., those in Figure~\ref{fig:triangle_SB}),
and translates them to C++ implementations 
embedded with stream ISA assembly instructions.

One major challenge is stream management during code generation (similar to register allocation in traditional compilers). 
To implement an intersection, 
the compiler may generate instructions that introduce up to three active streams---two input streams loaded by \texttt{S\_READ} and one output stream produced by \texttt{S\_INTER}. 
We release these created streams eagerly, since resources used to maintain actives streams \blue{(e.g. s-cache and stream registers)} are limited.
The streams created by \texttt{S\_READ} are released by \texttt{S\_FREE} after the intersection operation, 
\blue{and the compiler will insert \texttt{S\_FREE} instructions to free the stream produced by \texttt{S\_INTER} once it is no longer needed.}
If the number of actives streams reaches its limit (i.e., the number of stream registers), 
the compiler simply falls back to generate scalar ISA based intersection code, 
and print outs a warning message. 
In practice, we notice that such a ``fall-back'' scenario is rare (and actually did not happen for all applications evaluated in this paper) thanks to our aggressive stream freeing strategy.

\projectname also provides the 
 low-level programming interface for the
 experienced users to construct more 
 complex applications such as frequent subgraph 
 mining (FSM).
Table~\ref{table:api} lists \projectname APIs for 
building GPM applications by users. 
Each function can be 
implemented with the proposed stream ISA extension. 
Figure~\ref{code:full_example} shows an example of {three-chain algorithm} based on the APIs.

\section{Evaluation}
\label{sec:eval}


\subsection{Simulator and Configuration}
\label{sec:simulator}
We simulate \projectname on zSim \cite{Sanchez-zsim}, a fast and scalable simulator designed for x86-64 multicores. 
We integrate all our proposed architectural components 
including Stream Cache and Intersection Unit into the simulator. 
Our configuration is listed in Table \ref{table:config}. 

\begin{table}[]
\centering
\begin{tabular}{l|l}
\hline
Number of cores  &  8  \\ \hline
ROB size &  128   \\ \hline
loadQueue size &  32   \\ \hline
l1d cache size &  64KB   \\ \hline
cache line size & 64B \\ \hline
l2(last level) cache size  &   2MB\\\hline
stream cache latency  &   1 cycle\\\hline
stream cache bandwidth  &   16B\\\hline
stream cache slot size  &  256B \\\hline
l1d latency  &   4 cycles\\\hline
l2(last level) cache latency  &   10 cycles\\\hline
memory controllers number  &   3\\\hline
memory controller latency  &   40 cycles\\\hline
memory type  &   DDR3-1333-CL10\\\hline
\end{tabular}
	\caption{\projectname Architecture Configuration}
	\label{table:config}
		\vspace{-3mm}
\end{table}


\subsection{Graph Mining Algorithms and Data Sets}
\label{sec:algo_data}

\begin{table}[]
\scalebox{0.75}{
\begin{tabular}{l|l|l|l|l}
 \hline
name & \#V & \#E & avg D & max D\\ \hline
citeseer   (C) \cite{bader2013graph,cs_graph,nr-aaai15}  &  3.3K   &   4.5K & 1.39 & 99\\ \hline
email-eu-core   (E)\cite{leskovec2007graph,yin2017local}  & 1.0K    &  16.1K & 25.4 &345\\ \hline
soc-sign-bitcoinalpha  (B) \cite{konect:2018:soc-sign-bitcoinalpha,konect:kumar2016wsn,konect}  &   3.8K  &24K & 6.4&511\\ \hline 
p2p-Gnutella08  (G) \cite{p2p-G1,p2p-G2}  &  6k   &  21k & 3.3 &97\\ \hline
socfb-Haverford76  (F) \cite{nr-aaai15}  &  1.4K   &   60K & 41.3&375\\ \hline 
wiki-vote  (W)  \cite{leskovec2010signed,leskovec2010predicting} &   7k  &  104k & 14.6&1065\\ \hline 
mico  (M)  \cite{elseidy2014grami} &   96.6K  &  1.1M  &11.2&1359\\ \hline 
com-youtube  (Y) \cite{yang2012defining}   &   1.1M  &  3.0M &2.6&28754 \\ \hline 
patent  (P) \cite{leskovec2005graphs}  &   3.8M  & 16.5M & 8.8 &793 \\ \hline 
livejournal  (L) \cite{backstrom2006group,leskovec2009community}   &  4.8M   &  42.9M & 17.7&20333\\ \hline 
\end{tabular}
}
	\caption{Graph Datasets}
	\vspace{-7mm}
	\label{table:datset}
\end{table}

\setlength{\intextsep}{8pt}%
\setlength{\columnsep}{8pt}%
\begin{wraptable}[8]{r}{.5\linewidth}
	\centering
	\vspace{-2mm}
	\scalebox{0.8}{
	\begin{tabular}{|l|}
	\hline 
		name  \\ \hline
		Triangle counting    (T)  \\ \hline
		Three chain counting   (TC)   \\ \hline
		Tailed triangle counting  (TT)   \\  \hline
		3-motif  (TM)\\ \hline 
		4-clique  (4C)  \\ \hline 
		5-clique  (5C) \\ \hline 
		Frequent subgraph mining (FSM)\\ \hline
		
	\end{tabular}
	}
			\caption{Applications}
			\label{table:algorithms}
\end{wraptable}
We execute our  \blue{InHouseAutomine} ~\cite{mawhirter2019automine} to mine different 
patterns. 
We choose several popular GPM applications listed in Table~\ref{table:algorithms} to evaluate \projectname. They can be divided into four categories.
(1) \emph{Pattern counting} applications, which include triangle (T), three-chain (TC), and tailed-triangle counting (TT). These three workloads aim to count the number of
triangle/three-chain/tailed-triangle embeddings, respectively.
We use T, 4C, and 5C to denote the nested implementations while TS, 4CS, and 5CS refer to the corresponding stream implementations without nested support.
(2) \emph{${\bm k}$-motif mining}, which counts the embeddings of all connected patterns with a given size $k$. 
(3) \emph{${\bm k}$-clique mining}, which discovers all size-$k$ complete subgraphs of the input graph. 
(4) \emph{Frequent subgraph mining (FSM)}, which aims to discover all vertex-labeled frequent patterns. A pattern is considered as \textit{frequent} if and only if its \textit{support} is no less than a user-specified threshold. Pattern support could have different definitions. However, all of them should satisfy the \textit{Downward Closure Property}, which requires that one pattern should never have a greater support than its subpatterns. This key property is used in FSM to prune the searching space efficiently--if one pattern is infrequent, it can be safely discarded since it cannot be extended to any frequent patterns. 
Similar to previous systems like Peregrine~\cite{jamshidi2020peregrine}, we choose the \textit{minimum image-based support metric}~\cite{bringmann2008frequent} and only discover frequent patterns with no more than three edges.
Besides, it is worth noting that GRAMER {\em mistakenly} used the pattern count (i.e., the number of embeddings) as the pattern support for FSM, which violates the Downward Closure Property. {\em We also implement this incorrect FSM algorithm for performance comparison purposes and refer to it as simple-FSM (sFSM) in our experiments}. 
Table~\ref{table:datset} lists the real-world graphs we used from various domains, ranging from social network analysis to bioinformatics. 



\subsection{Overall Performance }
We compare \projectname with GRAMER~\cite{yao2020locality} and our CPU baseline on different datasets and algorithms.


\subsubsection{Comparison with GRAMER}
\begin{figure}[h]
    \centering
    \includegraphics[width=\linewidth]{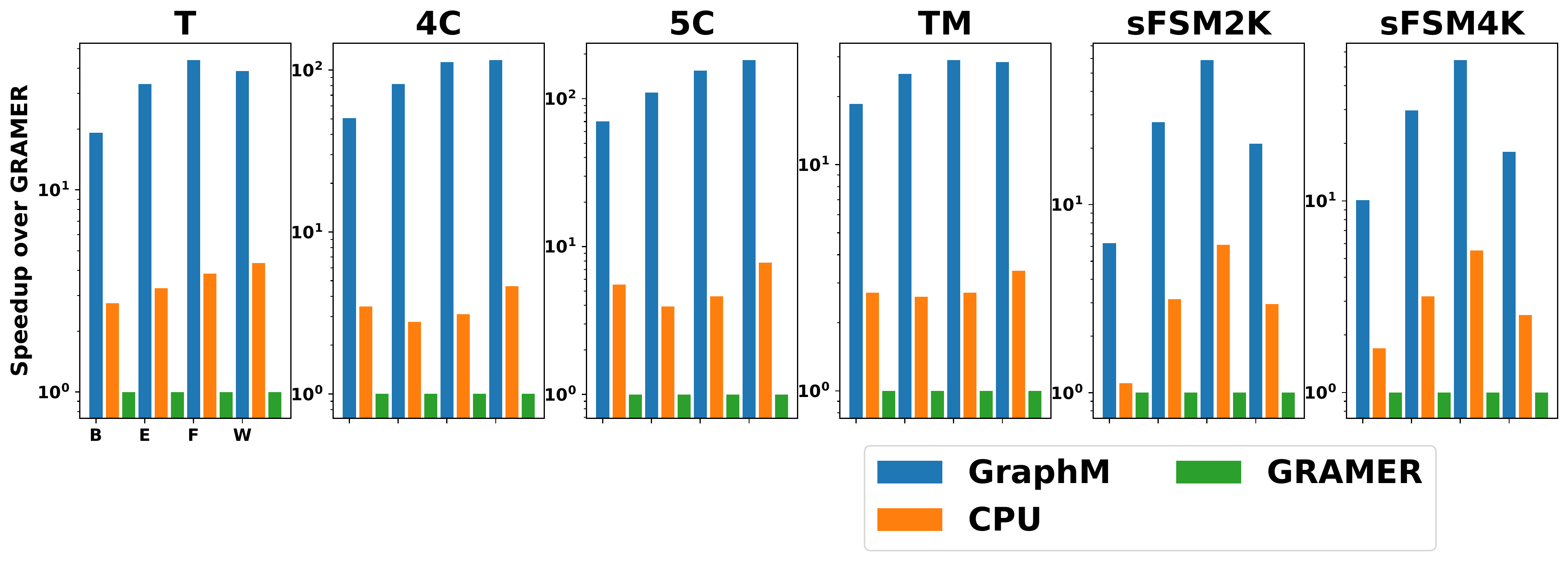}
    \vspace{-3mm}
    \caption{Speedup of \projectname and CPU (pattern enumeration) over GRAMER (exhaustive check) (log scale) 
    }
    \vspace{-4mm}
	\label{fig:cmp_gramer}
\end{figure}

We also implemented GRAMER on zsim. 
To simplify the comparison, we only enable one PU/CPU core in both GRAMER and \projectname.
In order to make a fair comparison, we configure GRAMER's on-chip RAM access latency to be the same as \projectname's first level cache latency. 


We compare the performance of \projectname, the CPU baseline, and GRAMER in Figure~\ref{fig:cmp_gramer}. The applications involved are Triangle Counting, 4/5-Clique, 3-motif, and sFSM with 2K/4K support thresholds. \projectname significantly outperforms both the CPU baseline and GRAMER. It is worth noting that GRAMER is even slower than our CPU baseline. The performance gap is majorly attributed to the algorithmic difference. Our CPU baseline implements the pattern enumeration method, which is much faster than the exhaustive check method in GRAMER. The architectural supports in GRAMER
cannot benefit the pattern enumeration method.

Another key observation is that \projectname achieves higher speedups over GRAMER for more complex patterns, such as 4/5-clique counting. On average, \projectname outperforms GRAMER by {\gramerFourCliqueAvgSpeedup, \gramerFiveCliqueAvgSpeedup} for 4-Clique and 5-Clique counting, respectively. By contrast, for triangle counting, the speedup is only \gramerTriAvgSpeedup. The speedup difference is reasonable. As we have discussed in Section~\ref{sec:Background}, one major source of exhaustive check method's inefficiency is their connectivity check operations.
A larger pattern incurs more connectivity checks. For instance, to extend a size-$k$ subgraph $(v_0,v_1,...,v_{k-1})$, the exhaustive check method typically selects an existing vertex $v_i(0\le i\le k-1)$, and choose one of its neighbor $v_k\in N(v_i)$ to be the new vertex. $k-1$ connectivity checks are needed to determine whether $v_k$ is connected with the existing vertices except for $v_i$. As a result, to enumerate a larger subgraph, exhaustive check systems/accelerators like GRAMER suffer more from the overhead caused by connectivity checks. This explains \projectname's higher speedup over GRAMER for the larger pattern since it is based on pattern enumeration and totally avoids connectivity checks.

\subsubsection{Comparison with CPU}
\label{sec:perf_cpu}
\begin{figure*}
    \centering
    \includegraphics[width=\linewidth]{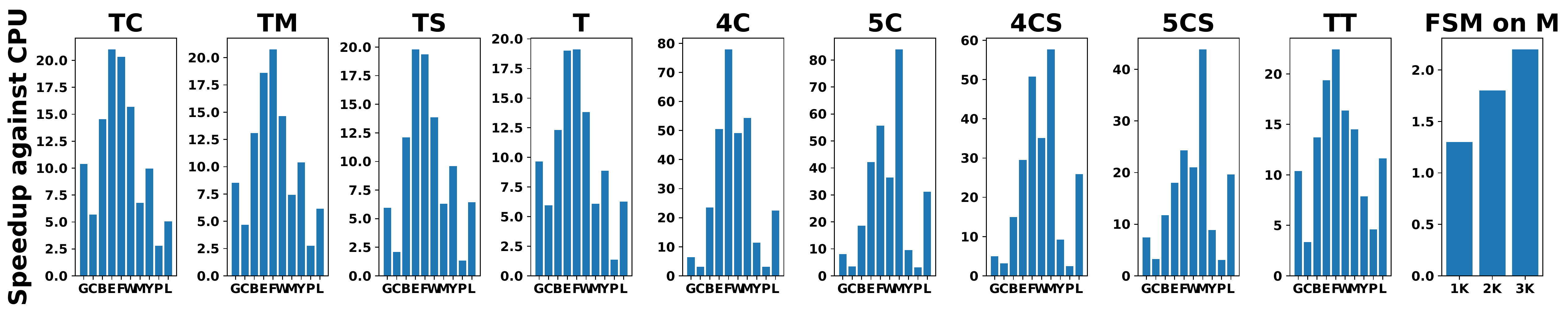}

    \caption{Speedups over CPU (Both use pattern enumeration)}
	    \vspace{-5mm}
	\label{fig:cmp_cpu}
\end{figure*}

Further performance comparison among \projectname (with/without nested intersection) and the CPU baseline are shown in Figure~\ref{fig:cmp_cpu}. TS, 4CS, and 5CS refer to the triangle counting, 4-clique, and 5-clique implementations without nested intersection. On average, enabling nested intersection speeds up these applications by \nestSpeedUp. 
It is because with nested intersection instructions, the normal instructions used to explicitly manage the corresponding loops, graph structure accesses, and embedding counting are eliminated. \blue{Nested intersection instructions allow more intersections execute on-the-fly simultaneously, thanks to the
reduction of normal instructions that would have occupied more ROB entries.} \red{Besides, note that \projectname achieves less speedup for FSM. It is because the support calculation in FSM is costly, and thus the intersection/subtraction operations that our architecture accelerates \blue{only} take a smaller portion of execution time.}



\label{nest_compare}

Comparing across different datasets, \projectname achieves higher speedups on graphs with higher average degree. 
This could be explained by Amdahl's law. 
On graphs with higher degrees, the operand lengths of intersection/subtraction operations are generally longer. As a result, these operations are more computation-intensive and take up a larger portion of execution time. Recall that \projectname only speedups intersection/subtraction operations, and thus achieves higher performance improvement on denser graphs.

\subsection{{Execution Time of Intersection Operations}}
\label{breakdown}

\begin{figure}[h]
    \centering
    \includegraphics[width=\linewidth]{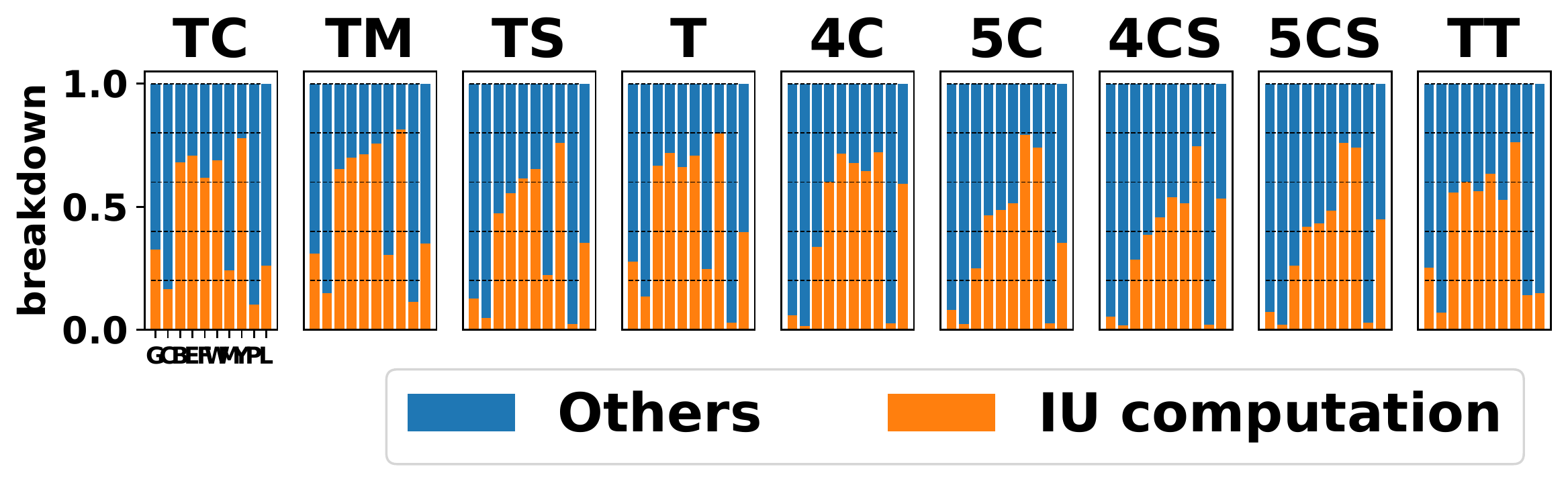}
    \caption{Execution Time Percentage of Intersections}
    \vspace{-6mm}
	\label{fig:ex_breakdown}
\end{figure}

We analyze the execution time percentage for intersection operations in \projectname. The results are shown in Figure~\ref{fig:ex_breakdown}. 
We observe that intersection operations still take up a large portion of execution time even though significant acceleration. This indicates that there are plenty of research opportunities to speedup GPM applications by optimizing intersections.

\subsection{\rev{Comparing to GPU}}
\label{perf_gpu}
\begin{figure*}[h]
    \centering
    \includegraphics[width=\linewidth]{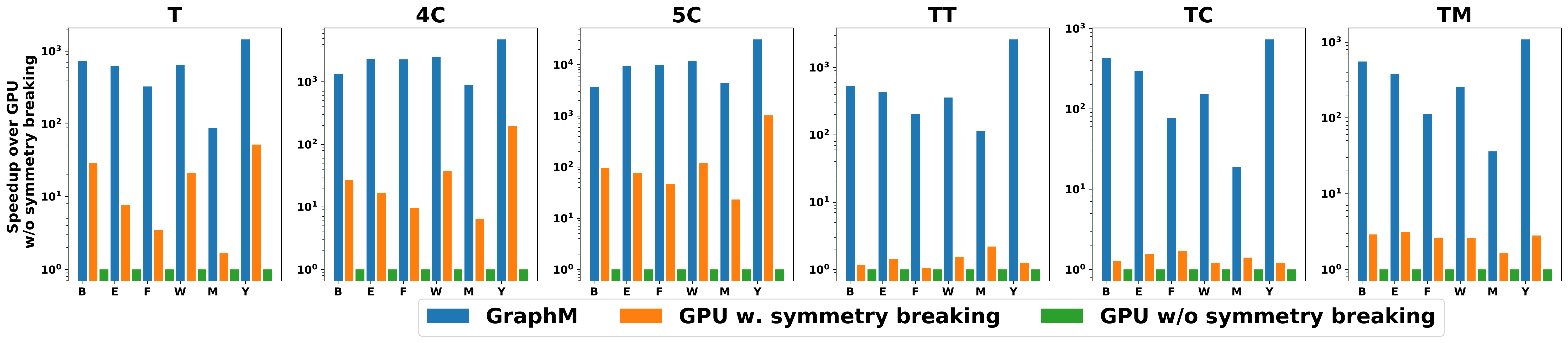}
    \caption{\rev{\projectname compared to GPU implementations (log scale)} }
    \vspace{-4mm}
	\label{fig:gpu}
\end{figure*}

\rev{We also compare \projectname with GPU (Nvidia Tesla K40m).
We assume the clock frequency of \projectname to be 1Ghz. 
We compare the performance of \projectname (with
symmetry breaking) with two GPU implementations
with or without symmetry breaking optimizations. 
The optimization in general \blue{adds} more branches,
and we want to study, with massive parallelism, whether 
the redundant enumeration with less branch divergence can overshadow less computation 
with more branches.
Figure~\ref{fig:gpu} shows the results. 
We can see that:
1) \projectname outperforms the GPU
implementations significantly, thus, even with
a more powerful GPU, the results should stay
the same; and 
2) symmetry breaking is also effective in GPU, and the massive parallelism on more computation cannot overweight less computation with more branches.
Using Nvidia profiling tools, 
we find that the reason for low performance of 
pattern enumeration on GPU is two-fold:
1) low warp utilization (about 4.4\%) due to the branches and 
the different loop sizes (edge list length) for different threads; and 2) low global memory bandwidth utilization (about 13\%)
since threads access edge 
lists at different memory locations. 
Based on our results, it is no surprise that all
existing pattern enumeration based graph 
mining system are based on CPU. }

\begin{figure*}[h]
    \centering
    \includegraphics[width=\linewidth]{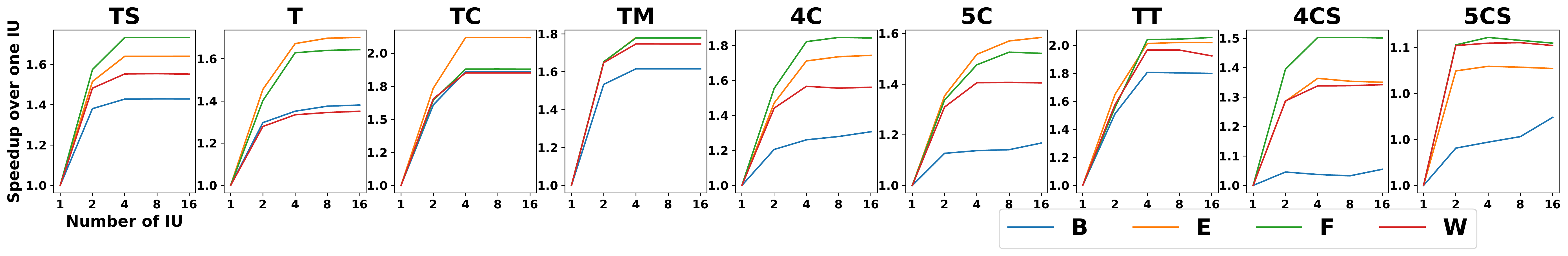}
    \vspace{-7mm}
    \caption{Varying the Number of IUs}
        \vspace{-5mm}
	\label{fig:sens_iu}
\end{figure*}
\begin{figure*}[ht!]
    \centering
    \includegraphics[width=\linewidth]{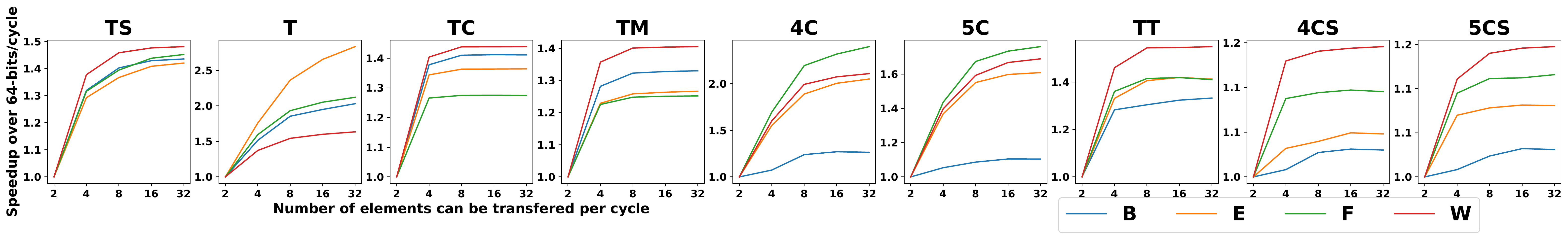}
    \vspace{-5mm}
    \caption{\rev{Varying S-Cache Bandwidth}} 
    \vspace{-6mm}
	\label{fig:bandwidth}
\end{figure*}

\subsection{\red{The Distribution of Stream Lengths}}

\begin{figure}[h]
    \centering
    \includegraphics[width=\linewidth]{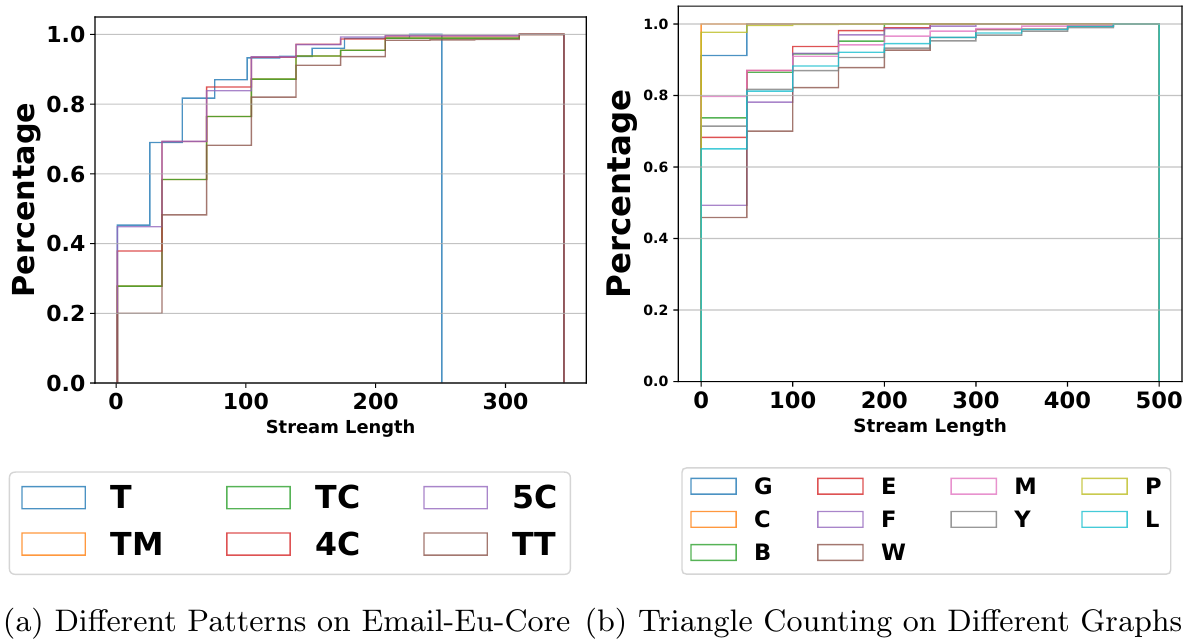}
    \caption{
    The Stream Length Distribution
    }
	\label{fig:CDF}
\end{figure}

We further analyze the length distribution of involved streams in different GPM algorithms. 
Figure~\ref{fig:CDF} (a) shows the cumulative distribution function (CDF) of stream lengths in different graph mining algorithms on the email-eu-core graph. 
Even on the same graph dataset, different applications could lead to different stream length distributions. 
We notice that clique applications (i.e., 4-clique/5-clique counting) in general introduce shorter stream lengths. 
The reason is that in clique applications, the input operands of intersection operations are usually the intersection \blue{results of other} streams. And these operands tend to have \blue{shorter} stream lengths. 

We also fix the graph mining application to triangle counting and analyze the stream length distribution on various datasets. The results are reported in Figure~\ref{fig:CDF} (b). For this figure, we cut off the counting for stream larger than 500. The observation is intuitive--the longest stream length on datasets with larger maximal degrees (e.g., LiveJournal, Youtube) are longer. Besides, there are more long streams on denser datasets like E (email-eu-core) and F (socfb-Haverford76).


\subsection{Varying the Number of Intersection Units}

We characterize the performance of \projectname by varying the number of IUs.
Figure \ref{fig:sens_iu} shows the results with 1 to 16 intersection units.
When the number of IUs is no more than 4, increasing it will generally improve \projectname's performance. However, with more than 4 intersection units, adding IUs introduces significantly less benefit or even slight performance degeneration. 
\blue{The non-monotonic behavior when increasing the numbers of IUs can be explained as follows. We find that the miss rate in L2 cache is increased in these cases. We believe this is due to the different IUs trying to read streams from different memory addresses, which leads to relatively more random requests to L2. Thus increasing IU number may lead to higher conflict in LLC.}
This indicates that the different architecture components 
need to be matched based on performance characterization.

\subsection{Analysis on S-Cache Bandwidth}

We further characterize \projectname's performance with different S-Cache bandwidths.
Figure \ref{fig:bandwidth} shows the performance of \projectname with S-Cache bandwidth varying from 2 elements (64 bits) per cycle to 32 elements (1024 bits) per cycle.
In general, increasing S-Cache bandwidth can improve \projectname's performance. However, 
there is a point of diminishing return. For example, for the TC (three-chain counting) application, increasing the bandwidth from 4 to 32 elements/cycle introduces almost no benefit. It is because there are not enough concurrent active stream intersection/subtraction operations to saturate the S-Cache bandwidth. The number of concurrent active stream operations is determined by the application and implementation. Triangle counting (T) and 4/5-Clique (4/5C) counting use the nested intersection instruction to trigger intersection operations in a bursty manner, and thus there are more simultaneously on-the-fly intersections. Hence, T/4C/5C benefits more from S-Cache bandwidth increase than other algorithm/implementation without the nested instruction (e.g., 4CS, 5CS). \blue{Moreover, each algorithm has a unique intersection pattern, which leads to different number of simultaneously on-the-fly intersections. Therefore, each algorithm would benefit from the S-Cache bandwidth increase differently.}


\subsection{Tensor Computation}
\begin{figure}
    \centering
    \vspace{-3mm}
    \includegraphics[width=0.7\linewidth]{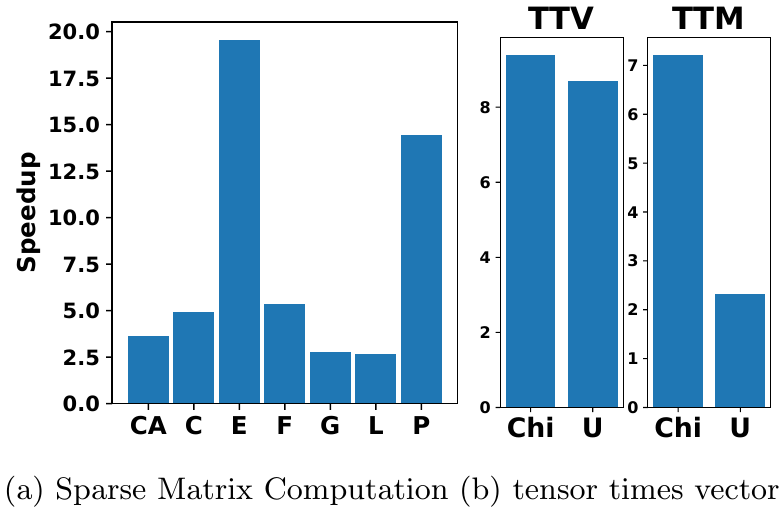}
    \vspace{-0mm}
    \caption{(key,value) Stream Computation Speedup}
    \vspace{-3mm}
	\label{fig:ky_speedup}
\end{figure}


To evaluate Stream Value Processing Unit
(SVPU) of \projectname, 
we implemented sparse matrix multiplication, \blue{tensor times vector (TTV, $Z_{ij} = \sum_k A_{ijk} B_k$) and tensor times matrix (TTM, $Z_{ijk} = \sum_l A_{ijl} B_{kl} $)} with our (key,value) stream interfaces.
For sparse matrix multiplication, we convert the matrix format from Compressed Sparse Row (CSR) to Compressed Sparse Column(CSC) and multiply the column of CSR with the row of CSC.
\rev{For tensor times vector and tensor times matrix, we store the tensor in CSF\cite{10.1145/2833179.2833183} format and the vector as sparse vector. We use state-of-the-art tensor algebra compilers (TACO\cite{kjolstad:2017:tacotool}) to generate baseline tensor kernel.}

We evaluated \projectname with matrices and tensors listed in Table~\ref{table:matrix}.
The speedup of \projectname against the CPU baseline is shown in Figure~\ref{fig:ky_speedup}.
For sparse matrix multiplication, \projectname achieves on average 5.7$\times$ speedup.
For matrices with higher density, \projectname can achieve higher speedup. The reason is, since there are more non-zeros, denser matrices lead to more intersection computation, which can be accelerated by \projectname architecture.
\blue{For tensor computation, \projectname achieves on average \red{6.08}$\times$ speedup. Similar to matrices, for tensor with higher density, \projectname can achieve higher speedup.}


\begin{table}[]
	\centering
\resizebox{.49\textwidth}{!}{	\begin{tabular}{l|l|l|l}
	\hline
	Name  & Dimensions &Nonzeros & \rev{Sparsity}\\
	\hline
	Circuit204  (C)\cite{davis2011university} & $1020 \times 1020$ &5883& 0.57\%\\
	Email-Eu-core(E)\cite{leskovec2007graph,yin2017local}& $1005\times 1005 $&25571&2.5\%\\
	Fpga\_dcop\_26 (F)\cite{davis2011university}&$1220   \times1220$ &5892 & 0.40\%\\ 
	Piston (P) \cite{davis2011university}&$2025 \times 2025$& 100015 & 2.4\%\\
	Laser (L)\cite{davis2011university}&$3002\times 3002$ &5000&0.055\% \\
	Grid2 (G)\cite{davis2011university}&$3296\times3296$ &6432 & 0.059\%\\
	Hydr1c (H) \cite{davis2011university}&$5308\times5308$& 23752& 0.084\%\\
	California (CA)	\cite{kleinberg1998authoritative,langville2006reordering}&$ 9664 \times 9664$ &16150 & 0.017\%  \\ 
	\hline
	\hline
		\rev{Chicago Crime} (Chi)\cite{frosttdataset}&$ 6186\times24\times2464 $ &5330673 & 1.46\%  \\ 
	\rev{Uber Pickups} (U)	\cite{frosttdataset}&$4392\times1140\times1717 $ &3309490 & 0.0385\%  \\ 
	\hline
	\end{tabular}}
	\caption{Matrix and tensor Datasets}
	\vspace{-6mm}
	\label{table:matrix}
\end{table}

\section{Conclusion}

This paper proposes \projectname, a  vertically
designed accelerator for GPM applications with
stream instruction set extension and 
architectural supports based on 
conventional processor. 
We develop the \projectname architecture composed of specialized mechanisms that efficiently
implement the stream ISA extensions.
We develop the \projectname architecture composed of specialized mechanisms that efficiently
implement the stream ISA extensions.
We implement \projectname ISA and 
architecture on zsim \cite{Sanchez-zsim}.
We use {7} popular GPM applications (\red{triangle/three-chain/tailed-traingle counting, 3-motif mining, 4/5-clique counting, and FSM}) on {10} real
graphs.
\red{Our extensive experiments show that \projectname outperforms our CPU baseline and GRAMER significantly. The average speedups could be up to \MaxSpeedupCPU and \MaxSpeedupGRMER and on average \AvgSpeedupCPU and \AvgSpeedupGRMER, respectively.}

\bibliographystyle{IEEEtranS}
\bibliography{references}

\end{document}